\documentclass{article} %
\usepackage[utf8]{inputenc}
\usepackage[margin=1.5in,top=1.0in,bottom=0.75in]{geometry}
\usepackage[english]{babel}
\usepackage{amsfonts}
\usepackage{amsmath}
\usepackage{multibib}
\usepackage{graphicx}
\newcites{si}{SI References}

\usepackage[super,sort&compress]{natbib}
\usepackage{natbib}
\usepackage[convert-footnotes=true,use-sort-key=false]{notes2bib}
\bibnotesetup{
  note-name = ,
  use-sort-key = false
}
\usepackage{longtable}
\usepackage{tabularx}
\usepackage{setspace}
\usepackage{url}
\usepackage{afterpage}
\usepackage{bm}

\usepackage{fancyhdr}

\newcommand{\mtsectionhead}[1]{\par\medskip\noindent\textbf{#1}\medskip\par}

\begin{document}

{\centering
{\Large Measuring frequency-dependent selection in culture}\par\medskip\par
 Mitchell G. Newberry$^1$, Joshua B. Plotkin$^2$\par\medskip\par
  {\footnotesize $^1$ Center for the Study of Complex Systems, University of Michigan 48109\\
    $^2$ Departments of Biology and Mathematics, University of Pennsylvania, 19104}
\par}

\setlength{\unitlength}{1pt}
\newcommand{\solidcircle}{\begin{picture}(6,6)(0,0)\put(3,3){\circle*{3}}\end{picture}}
\newcommand{\opencircle}{\begin{picture}(6,6)(0,0)\put(3,3){\circle{3}}\end{picture}}

\renewcommand{\abstractname}{} \begin{abstract}
\noindent Cultural traits such as words\citep{trudgill2004new},
names\citep{hahn2003drift}, decorative styles\citep{neiman1995stylistic}, and
technical standards\citep{wagner2016primordial} often assume arbitrary values
and are thought to evolve
neutrally\citep{kimura1983neutral,hahn2003drift,bentley2004random,neiman1995stylistic,reali2010words}.
But neutral evolution cannot explain\citep{kandler2013non} why some traits come
and go in cycles of popularity\citep{ghirlanda2013fashion} while others become
entrenched\citep{wagner2016primordial,pagel2019dominant}.  Here we study
frequency-dependent selection---where a trait's tendency to be copied depends
on its current frequency regardless of the trait value itself.  We develop a
maximum-likelihood method to infer the precise form of frequency-dependent
selection from time series of trait abundance, and we apply the method to data
on baby names and pet dog breeds over the last century.  We find that the most
common names tend to decline by 2\%-6\% per year on average; whereas rare
names---1 in 10,000 births---tend to increase by 1\%-3\% per year.  This
specific form of negative frequency dependence explains patterns of
diversity\cite{odwyer2017inferring} and replicates across the United States,
France, Norway and the Netherlands, despite cultural, linguistic and
demographic variation.  We infer a fixed fitness offset between male and female
names that implies different rates of innovation.  We also find a strong
selective advantage for biblical names in every frequency class, which explains
their predominance among the most common names. In purebred dog registrations
we infer a form of negative frequency dependence that is consistent with a
preference for novelty, in which each year's newest breeds outgrow the previous
by about 1\%/year, which also recapitulates boom-bust cycles in dog
fanciers\cite{ghirlanda2013fashion}.  Finally, we define the concept of
\textit{effective frequency-dependent selection}, which enables a meaningful
interpretation of inferred frequency dependence even for complex mechanisms of
evolution.  Our analysis generalizes neutral evolution to incorporate pressures
of conformity and anti-conformity as fundamental forces in social evolution,
and our inference procedure provides a quantitative account of how these forces
operate within and across cultures.
\end{abstract}

Humans adopt millions of different first names, hundreds of different dog
breeds, and virtually one word for baseball---even though names\cite{hahn2003drift}, dog
breeds\cite{ghirlanda2013fashion}, and words\cite{pagel2007frequency} are all
chosen from arbitrary alternatives. Why do skirt lengths continuously
fluctuate\cite{curran1999analysis} while media
formats\citep{katz1986technology} enjoy long periods of stasis? What's in a
name, if a rose by any other name would smell as sweet?

Here we advance frequency-dependent selection to unify different domains of
evolution.  We develop a method to quantify the form of frequency dependence,
which has been difficult to observe in biological data.  Socially relevant
traits in humans evolve by cultural transmission, through mechanisms of
imitation, learning, and innovation\cite{boyd1985culture,mesoudi2011cultural}.
Nonetheless, models of biological evolution that describe the dynamics of types
in reproducing populations\cite{ewens2012mathematical} can be used to describe
the dynamics of social traits inherited by cultural
transmission\cite{cavalli1981cultural,henrich2003evolution,henrich2004cultural,reali2010words}.
The neutral model of evolution\cite{kimura1983neutral} attributes changes in a
population's composition to the accumulated effects of indiscriminate random
copying, which already recapitulates empirical aspects of
names\cite{hahn2003drift} and
words\cite{reali2010words,trudgill2004newdialect,newberry2017detecting}.  Here
we generalize the neutral model by introducing selection---adjusting the
propensity to copy different traits---in the case when selection depends solely
on a type's frequency in the population. This frequency-dependent selection
remains blind to any distinguishing features of the types themselves and so
constitutes an exchangeable model of evolution\cite{cannings1974latent}, which
is mathematically tractable despite high dimensionality.  This generalization
of neutrality is analogous to density-dependent
extensions\cite{volkov2005density} of neutral biodiversity theory in
ecology\cite{hubbell2001unified}, and it corresponds to coordination and
anti-coordination games in evolutionary game theory\cite{smith1982evolution}.
By incorporating selection, the model encompasses a spectrum of behavior from
rapid diversification\cite{aguilar2004high} to intractable entrenchment of a
single type\cite{lande1979effective,wagner2016primordial}.  The model is an
effective base case for social evolution, because it captures the selective
pressures that are common to all types.

Biologists seldom have the opportunity to measure precisely how selection
depends on frequency in the wild. Despite this lacuna, frequency-dependent
selection has been long-studied\cite{hori1993frequency,futuyma2009evolution}
and commonly invoked in explanations of genetic
diversity\cite{fisher1930genetical}, immune
escape\cite{levin1988frequency}, altruistic
behavior\cite{matessi1976conditions}, sex ratios\cite{fisher1930genetical},
mating preferences\cite{weatherhead1979offspring}, reproductive
timing\cite{janzen1976bamboos} and
speciation\cite{wright1941probability,lande1979effective}.

Cultural traits in humans, by contrast, ``fossilize'' ubiquitously, often
leaving complete records of the abundance of alternative types over time.
Cultural data therefore provide a rich source of empirical
patterns\cite{michel2011quantitative} that we can use to infer specific forms
of frequency dependence in cultural
evolution\cite{boyd1985culture,efferson2008conformists,bowles2009microeconomics},
and to provide a proof of concept for studying how social behavior governs
selection on genetic and morphological traits in biology.

\mtsectionhead{Evolutionary model and parameter inference}

\begin{figure}[t] %
\noindent
\hspace{-25mm}
\includegraphics[width=183mm]{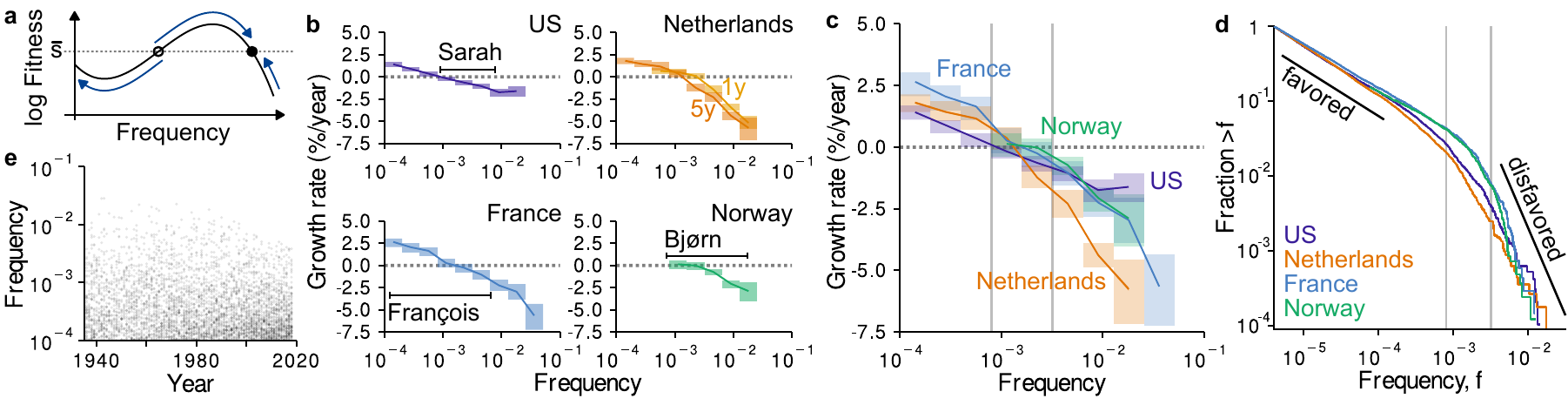}
\caption{\textbf{Frequency-dependent selection on first names.} Fitness is
frequency-dependent (a) when the growth rate of a trait depends on the trait's
frequency $p$ in the population. Traits with log-fitness $s(p)$ above the mean
selection coefficient $\bar{s}$ tend to increase in abundance, while those
below the mean fitness tend to decrease. We infer frequency-dependent selection
from time series of names given at birth (e: U.S.~Social Security
Administration database, 5\% sample).  A name's fitness depends on its current
frequency (b) in the United States, the Netherlands, France, and Norway, while
any particular name (e.g.~Sarah, François, Bjørn) may occupy different
frequency bins and growth rates from year to year.  Uncensored anonymized
counts from the Netherlands allow us to probe any affects arising from temporal
discretization, as well as study frequencies lower than than the annual birth
cohort size, by aggregating data into either 1-year or 5-year time intervals.
All four countries show similar patterns of negative frequency-dependence (c),
with an intercept at zero growth rate for names at frequencies between 8 and
32 births per 10,000 (vertical grey lines).  Selection favoring names more rare
than this intercept, and disfavoring names more common, corresponds to
different slopes in the distribution of name frequencies (d), which features
cusps near the points of zero growth. In all panels shaded boxes indicate
frequency bin boundaries and bootstrap 95\% confidence intervals
(\ref{sec:cis}).  Bins with fewer than 5 names or bias possibly exceeding
0.8\%/year are omitted.}
\label{fig:names}
\end{figure}

We model evolution as a Wright-Fisher process \cite{ewens2012mathematical}
where the growth rate $e^{s(p)}$ associated with a trait depends on the trait's
current frequency $p$ in the population, via a frequency-dependent selection
coefficient $s(p)$. The functional form of frequency-dependent selection,
$s(p)$, is assumed common to all traits, so that $s(p)$ completely specifies
the competitive environment of the exchangeable model (Fig.~\ref{fig:names}a).
The case $s(p) \equiv 0$ recovers neutral evolution, whereas $s(p)$ linearly
increasing or linearly decreasing with $p$ recover textbook formulations of
positive and negative frequency-dependent selection\cite{futuyma2009evolution}.
Mutations (innovations) occur at rate $\mu$ per generation and always produce
novel types.

We infer the form of frequency-dependent selection $\hat{s}(p)$ and the
mutation rate $\hat{\mu}$ from time series of trait frequencies, by maximum
likelihood (\ref{sec:sp}). We parameterize frequency dependence by assuming
$s(p)$ is a piecewise-constant function (Fig.~\ref{fig:pwconstx}), which
associates a selection coefficient $s_i$ with each frequency range (bin) $i$
with $p \in (l_i,u_i]$. This parameterization approximates a vast space of
possible frequency-dependent growth laws.  Maximizing likelihood gives
equations for the selection parameters $\hat{s}_i$ that are separable from the
mutation rate $\hat{\mu}$ and independent of population size, but inseparable
from each other so that no analytic solution for the $\hat{s}_i$ exists.
Instead, we iteratively approximate the maximum likelihood $\hat{s}_i$ to
arbitrary precision by optimizing a convex surrogate minorant
function\citep{lange2000optimization} (\ref{sec:sp}, code released publicly).
And so we obtain a nonparametric approximation of $s(p)$; more data allows
either finer frequency resolution at a given level of precision, or more
precise estimates of selection at a given resolution of frequency bins.

We compute confidence intervals on parameters by two methods---one derived
analytically from Fisher information and one derived empirically from an Efron
boostrap adapted for time series data (\ref{sec:cis}). We display only the more
conservative, bootstrap confidence intervals, because these are both wider and
less sensitive to model misspecification. We furthermore control for biases due
to censorship of rare types (\ref{sec:cen}) and due to sampling error
(\ref{sec:samp}) respectively by deriving analytic upper- and lower-bound
imputations of missing data and inferring $\hat{s}(p)$ from a synthetic
time series of samples from the static distribution of observed frequencies.
Both biases primarily affect low frequencies.  We also empirically estimate
bias due to discrete time interval data (\ref{sec:names},
Fig.~\ref{fig:names}b, Netherlands).

\mtsectionhead{Frequency-dependent selection in baby names}

Naming is a natural case study for frequency-dependent selection in cultural
evolution. Whereas Galton, Lotka, and Fisher studied inheritance of family
names as a model of extinction by stochastic
drift\cite{watson1875probability,lotka1931population,feller1951diffusion}, we
study first names. First names rarely have observable lines of inheritance, and
yet contemporary research has implicated
drift\cite{hahn2003drift,bentley2004random} as well as a diverse host of
psychological and social forces\cite{lieberson1992children,lieberson2000matter}
in the dynamics of first names. This rich catalog includes novelty
bias\cite{odwyer2017inferring}, immigration and assimilation
\cite{lieberson2000matter,goldstein2016patrick}, class imitation and aversion
\cite{lieberson2000matter,goldstein2016patrick}, trend momentum
\cite{gureckis2009you,berger2009adoption,kessler2012you}, phonological
affects\cite{lieberson2000matter,mutsukawa2011phonological,berger2012karen},
and cultural broadcasts\cite{lieberson2000matter}.

We examined frequency dependence in the dynamics of first names in the United
States (US) using the Social Security Administration baby name database, which
includes first name, birth year, and assigned sex of nearly all Social Security
Card recipients born in the United States. We inferred the frequency-dependent
growth curve $\hat{s}(p)$ (Fig.~\ref{fig:names}) from a time series beginning at
the administration's inception in 1935, using the 245 million individuals whose
names occur in at least 1 in 10,000 births (Fig.~\ref{fig:names}e). By omitting
names yet more rare, as well as any frequency bins with fewer than 5 names, we
remove noise and bias that would otherwise arise from sampling or censorship in
the US dataset (\ref{sec:names}). 

According to our maximum-likelihood inference, when a name is rare, at
frequency near 1 in 10,000 births, it has an intrinsic growth rate of
1.4\%/year whereas the most common names, above 1 in 100 births, decline at
-1.6\%/year (Fig.~\ref{fig:names}b).  Between these two extremes, the growth
rate of a name declines linearly with its log-frequency ($r^2$$=$0.98).  The
frequency of any particular name varies over time (such as ``Sarah'',
Fig.~\ref{fig:names}b), so the same name can be favored by selection at some
times and disfavored at others.

Our inference of frequency-dependent selection on names, $\hat{s}(p)$, easily
rejects neutrality, $s(p) \equiv 0$, with a $p$-value $<$0.0001, in contrast to
prior studies that could not reject neutrality from the stationary distribution
of name frequencies\cite{hahn2003drift,odwyer2017inferring}.

Negative frequency dependence reflects what it means to be a name: a name's
value derives predominantly from its uniqueness. For example, parents are known
to abandon names that are becoming too popular\cite{berger2009adoption}. In
aggregate, this and many other psychological and social
factors\cite{lieberson2000matter} repress a name when it is common and promote
it when rare.

\mtsectionhead{Cross-cultural patterns of name dynamics and diversity}

France, the Netherlands, and Norway also publish comprehensive time series of
first names (Table~\ref{tab:babstats}), and we infer similar patterns of
frequency-dependent selection despite the different names, naming conventions,
languages, cultures, demographics, and population sizes across these countries
(Fig.~\ref{fig:names}b-c, \ref{sec:names}).  The curve $\hat{s}(p)$ intersects
zero growth at a frequency that is consistent across countries---between 8 and
32 in 10,000 births (Fig.~\ref{fig:names}c)---to within a factor of four
(Fig.~\ref{fig:names}c), despite a nearly 70-fold difference in population
sizes (Table~\ref{tab:babstats}).  This means that different cultures have
similar sensitivities to what specific frequencies make a name desirable or
undesirable.

The frequency where selection transitions from favoring to disfavoring names is
further reflected in the stationary distribution of name frequencies---the
diversity of names\cite{hubbell2001unified,hahn2003drift}.  Plotting the
fraction of names that exceed a given frequency on log-log axes
(Fig.~\ref{fig:names}d) reveals two distinct regimes punctuated by a cusp near
the frequency of zero growth, across all four countries.  The diversity of
favored, rare names shows a clear power-law slope, whereas a steep cutoff
occurs among the common, disfavored names.  Prior studies have reported either
a power law with a single slope\cite{hahn2003drift} or predicted two
regimes\cite{odwyer2017inferring}, whereas here we confirm two distinct regimes
that furthermore correspond to our inference of whether names are favored or
disfavored by selection. Our inference uses time series data and a dynamical
model, yet it explains general features of the frequency distribution of names
averaged over time.

\mtsectionhead{Gendered and biblical names: growth laws within subpopulations}

\begin{figure}[t]
\noindent
\centering
\includegraphics[width=120mm]{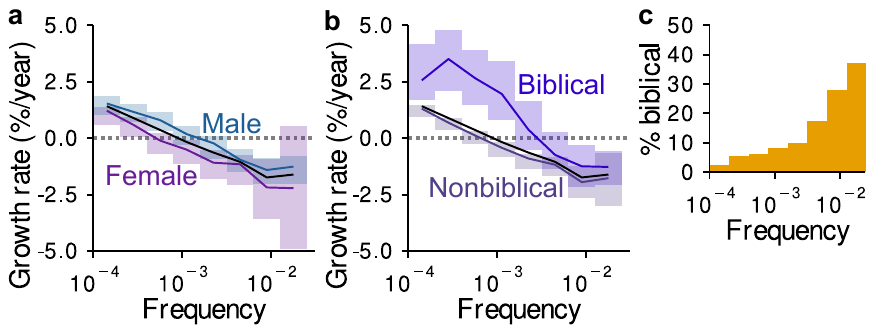}
\caption[]{\textbf{Growth laws for female, male, biblical, and non-biblical
names in the United States.} Frequency-dependent fitness differs between the
subpopulations of names given to females and males (a) and between biblical
and non-biblical names (b), with higher fitness among male and biblical names.
The inferred frequency-dependent fitness for the entire population (black) is
the weighted average fitness of its subpopulations. Lower growth rates for
female names across all frequencies is consistent with dilution caused by
higher rates of innovation among female names. Biblical names, by contrast,
preclude innovation and they avoid dilution through higher intrinsic fitness,
as evidenced by their inferred frequency-dependent growth law (b) and the
enrichment of biblical names among top names (c).}
\label{fig:mfbib}
\end{figure}

Names given to males or females represent subpopulations that may differ in
their growth rates and forms of
frequency-dependence\cite{lieberson2000matter,lieberson1992children}.  By
stipulation, however, our exchangeable model treats all types in the population
uniformly. A modest extension to the model allows us to infer a
frequency-dependent growth curve within a subpopulation of types
(\ref{sec:subpop}). We simultaneously infer a single, time-independent
``wildtype'' fitness for the remainder of the population, which calibrates the
zero point of growth rate (\ref{sec:wildtype}). Separate inferences for
mutually-exclusive subpopulations produce frequency-dependent growth laws for
each subpopulation that are consistent with the growth law measured for the
whole population as the weighted average of subpopulations. This protocol
accurately reconstructs the form of frequency-dependent selection in
subpopulations, even when subpopulations have radically different growth laws
(Fig.~\ref{fig:parlin}).

Separating male and female persons in the US dataset---treating for example
``Frances,M'' and ``Frances,F'' as distinct names in different
subpopulations---we find an appreciable difference in growth rates
($\sim$0.66\%/year, Fig~\ref{fig:mfbib}a).  This fitness difference between
male and female names is roughly constant across the full range of name
frequencies (Pearson $\rho$$=$0.40, $p$$=$0.33).  Of course the total number of
male names cannot increase relative to female names over time because the ratio
of female to male persons is constrained to roughly 1:1. And so the constant
difference in growth rates between male and female names must be compensated by
different rates of innovation: the lower growth rates of extant female names
must be balanced by a greater influx and higher turnover of novel female names.
The rate of innovation (mutation) affects what selection value corresponds to
zero net growth (Eq.~\ref{eq:rep} in \ref{sec:rep}) and hence displaces $s(p)$
uniformly across frequencies, as observed in Fig~\ref{fig:mfbib}a. 

There are 1.63 female names per male name in the dataset, compared to 0.98
female persons per male person. A range of mechanisms have been proposed to
explain the greater diversity of female names, including a more diverse
standing repertoire for recombination and spelling
variation\cite{lieberson1992children,lieberson2000matter}, greater sensitivity
to fads\cite{berger2009adoption} and gender-specific tastes in etymology and
phonology\cite{barry2010racial}.  Our result suggests that different innovation
processes, rather than different dynamics, explain the elevated diversity of
female names.

Biblical and non-biblical names also represent distinct subpopulations of
names, and there are striking patterns in their overall frequencies over
time\cite{gerhards2000trends,lieberson2000matter}.  Whereas some studies
attribute the dynamics of biblical names to variation in religiosity
\textit{per se}\cite{gerhards2000trends}, associated for example with
Catholicism\cite{perl2004dont}, others conclude that biblical names are subject
to the same social pressures as the population of names as a
whole\cite{lieberson2000matter}.

We infer that biblical names also experience negative frequency-dependent
selection, and yet they enjoy substantially greater fitness than non-biblical
names, across all frequencies (Fig.~\ref{fig:mfbib}b).  Since selection favors
biblical names regardless of their frequency, we should expect to find
enrichment for biblical names among the top names. Indeed we observe this
enrichment: biblical names account for nearly 40\% of names in the highest
frequency bin compared to only 2\% of names in the lowest frequency bin
(Fig.~\ref{fig:mfbib}c).  Enrichment also explains the reduced fitness
advantage of biblical names relative to the whole population at higher
frequencies (Fig.~\ref{fig:mfbib}b), simply because the whole population
includes those high fitness biblical names.

By extension, any idiosyncratic advantages---including biblical meaning or
phenology\cite{berger2012karen}---are especially likely to predominate among
names at the highest frequencies. And so, because our exchangeable model
ignores idiosyncratic selective benefits, inferred selection among the top
names represents a combination of the frequency-dependent \textit{dis}advantage
to any common name and frequency-\textit{in}dependent advantage associated with
those particular names.

\mtsectionhead{Frequency-dependent selection and novelty bias in dog breed
preference}

\begin{figure}[t]
\noindent 
\centering
\includegraphics[width=110mm]{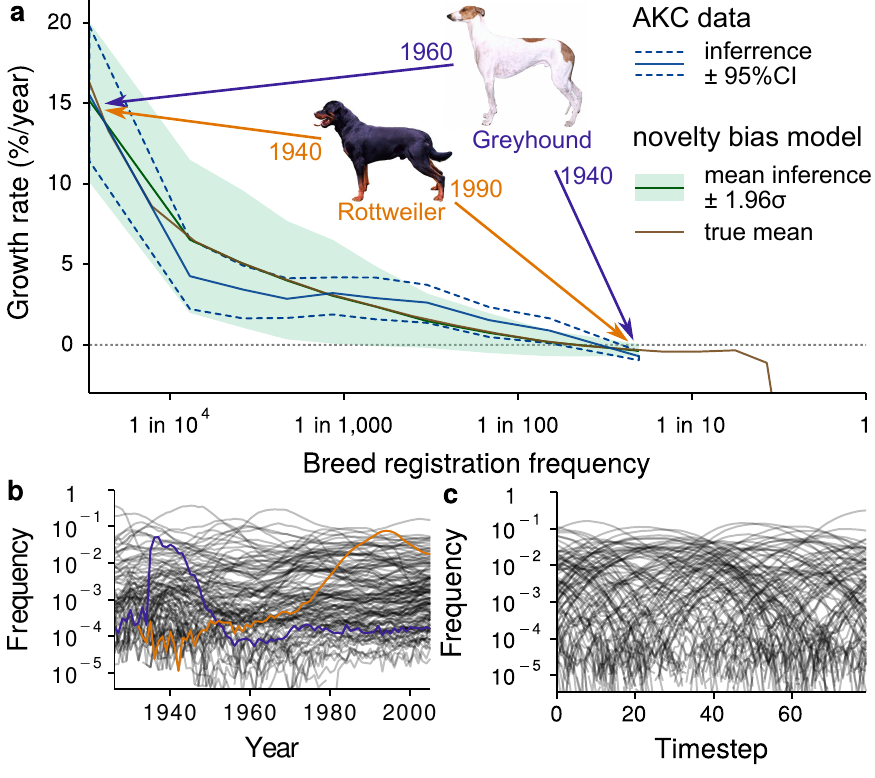} 
\caption{\textbf{Frequency dependence in dog breeds} We infer negative
frequency-dependent growth (a, blue) in a time series (b) of purebred dog
registrations in the American Kennel Club (AKC), comprising 153 distinct breeds
over 80 years. The rarest breed registrations increase on average by 16\%/year
while the commonest slowly decline.  Popular dog breeds are subject to fads:
several breeds, such as Greyhound (a-b, purple) and Rottweiler (a-b, orange),
cover the entire frequency range. Simulations of a novelty bias model display
similar boom-bust cycles (c) and match the frequency dependence inferred from
the AKC data (a, green) with parameters $N\mu\Delta{s}$$=$0.01, representing an
annual rate of selective improvement of 1\%/year. Note that mean
frequency-dependent selection $\hat{s}(p)$ inferred by maximum likelihood from
500 replicate simulations of the novelty bias model (a, green) overlaps with
the true average frequency-dependent selection $\bar{s}_i$ (\ref{sec:avg}) of
the novelty bias model (a, brown).  Novelty bias parameters (a,c):
$\mu$=0.00004, $\Delta s$=0.00021, $N$=1,435,737, $N_e$=50,000.}
\label{fig:dogs}
\end{figure} 

Preferences---from movie genres to fashion trends---are among the earliest
case studies of cultural evolution\cite{cavalli1981cultural}.  In particular,
breed preferences among purebred dog collectors follow trends in popular
culture \cite{ghirlanda2014dog,herzog2004random}, even to the exclusion of the
health or behavioral attributes of the dogs
themselves\cite{ghirlanda2013fashion}. And so the breed preference of a
collector can be studied as a cultural
trait\cite{ghirlanda2014dog,herzog2004random,ghirlanda2013fashion}.

We infer the form of frequency-dependent selection $\hat{s}(p)$ in dog breed
preferences (Fig.~\ref{fig:dogs}a) from a complete annual time series of
purebred dog registrations with the American Kennel Club (AKC) from 1926 to
2005\cite{ghirlanda2013american} encompassing over 50 million dogs
(Fig.~\ref{fig:dogs}b, \ref{sec:dogs}). We find strong negative frequency
dependence---that rare breeds have higher growth rates on average than common
ones (Fig.~\ref{fig:dogs}a)---indicating that the diversity of breeds is
maintained by selection rather than drift and innovation
alone\cite{herzog2004random}.

Yet selection on frequency \textit{per se} cannot explain boom-bust cycles. The
predictable pattern of rise and fall in breed preferences
(Fig.~\ref{fig:dogs}b) implies that a type has different fitnesses for the same
frequency at different times, which is incompatible with fitness determined by
frequency alone. More generally, dynamics of cultural traits including names
often feature boom-bust cycles\cite{acerbi2012logic,herzog2006forty} and trend
momentum\cite{stadler2016momentum} that have been attributed to a preference
for novelty\cite{odwyer2017inferring} or rapid abandonment of recent
fads\cite{berger2009adoption}. 

Rather, we illustrate that frequency dependence and boom-bust cycles can both
arise as joint consequences of an underlying mechanism of novelty bias.  We
simulate selection for novelty---a fundamentally non-exchangeable
process---using a Wright-Fisher model in which each novel type enters the
population at rate $\mu$ per capita per generation, and it has a yet greater
fitness $\Delta s$ than the previous type (Fig.~\ref{fig:dogs}c).  We then
infer frequency-dependent selection $\hat{s}(p)$ from the simulation output and
we choose parameters so that the inference from the output matches the
inference from the AKC data (\ref{sec:dognov}).  The resulting novelty bias
simulations produce boom-bust cycles (Fig~\ref{fig:dogs}c) with timescale and
amplitude similar to those in the AKC data (Fig~\ref{fig:dogs}b).  Adjusting
the parameters $\mu$ and $\Delta s$ while fixing the population size $N$ and
drift rate $1/N_e$ to match the AKC data, we find good agreement between the
two inferences (Fig.~\ref{fig:dogs}a) whenever $N\mu\Delta s$---the average
increase in selective benefit per year of novelty---is 0.01 (CI: 0.004-0.036
caused by uncertainty in $N_e$).  In other words, collectors like new dogs but
the novelty wears off---by about 1\%/year each year.  

\mtsectionhead{Effective frequency-dependence}

The dynamics of human culture or biological evolution are far too complex to be
circumscribed by a single mathematical formulation. Indeed, in all the cases we
investigate, we do not propose that selection is determined by frequency alone,
as it is in the exchangeable model. Nonetheless, many complex mechanisms of
evolution, such as novelty bias, associate selection with frequency well enough
that biologists often conflate them with frequency-dependent
selection\cite{brisson2018negative}.

We tease apart this conflation by introducing the concept of \textit{effective}
frequency-dependent selection. We define effective frequency-dependent
selection---much like effective population size\cite{wright1931evolution}---as
the form of frequency-dependent selection that best matches the idealised
(exchangeable) population dynamics to the true (complex, non-exchangeable)
population dynamics.  For example, the novelty bias model induces a
characteristic effective frequency dependence (Fig~\ref{fig:dogs}a) even though
the model is not itself exchangeable.

In a case of mathematical serendipity, our inference procedure measures
effective frequency dependence for any evolutionary process.  Specifically,
frequency dependent selection $\hat{s}(p)$ inferred by maximum likelihood under
the exchangeable model converges to the true, average selection pressure
$\bar{s}_i$ on types within a given frequency range $p \in (l_i,u_i]$
(Fig.~\ref{fig:dogs}a, \ref{sec:avg}).  This correspondence holds for any
underlying selection mechanism, exchangeable or otherwise, because the
maximum-likelihood selection coefficients (Eq.~\ref{eq:fdml}) equal the
observed average selection within frequency ranges.  This result is
serendipitous because typically there is little to be gleaned from parameters
inferred under a misspecified model\cite{white1982maximum}. But in our case,
inferred frequency dependence is meaningful regardless.

Furthermore, effective frequency dependence is ``unreasonably
effective''\cite{wigner1990unreasonable} in describing models far beyond its
scope. For example, we fit parameters of the novelty bias model by matching
only exchangeable properties of the empirical data---effective frequency
dependence---and yet the chosen parameters also recapitulate the
characteristically non-exchangeable boom-bust cycles.  And so, effective
frequency dependence retains sufficient information to discriminate amongst the
vast space of all complex evolutionary processes.  

\mtsectionhead{Discussion}

Frequency-dependent selection offers an exchangeable lens on complex population
dynamics, by ignoring differences between types.  The diversity of social and
psychological forces that drive cultural evolution has stimulated an equally
diverse collection of
models\cite{hahn2003drift,bentley2004random,ghirlanda2013american,kandler2013non,henrich2004cultural,trudgill2004new,efferson2008conformists,gureckis2009you,berger2009adoption,kessler2012you,ghirlanda2014dog,acerbi2012logic,herzog2006forty,stadler2016momentum},
often developed in isolation. This rich library brings specificity and realism,
while the exchangeable lens provides a common currency for comparison and
calibration across domains of culture.

Although here we study culture, what we highlight is an important universal
aspect of population dynamics.  As genetic and phenotypic time series in biology
are increasingly available\cite{good2017dynamics}, inferring frequency
dependence in biological contexts can elucidate intraspecific
social\cite{west1979sexual} and interspecific
ecological\cite{connell1971role,janzen1970herbivores} processes underlying
evolutionary change. Pursuing this program within the framework of exchangeable
models will enable a broad consilience across domains of evolutionary theory,
from biology to culture.

\medskip

\noindent \textbf{Acknowledgement.} We thank Gerrit Bloothooft and Meertens
Instituut Nederlandse Voornamenbank for preparing anonymized first name data.

\bigskip

\bibliographystyle{unsrtnat}
\bibliography{index}

\clearpage

\setcounter{section}{0}
\renewcommand{\thesection}{S\arabic{section}}
\setcounter{page}{1}
\renewcommand{\thepage}{S\arabic{page}}

\noindent{\Large \textbf{Supplementary Information}}

\par\medskip\par

\setcounter{figure}{0}
\renewcommand{\thefigure}{S\arabic{figure}}

\setcounter{table}{0}
\renewcommand{\thetable}{S\arabic{table}}

\setcounter{equation}{0}
\renewcommand{\theequation}{S\arabic{equation}}

\tableofcontents

\section{Inference} 
\subsection{Likelihood maximization}

\label{sec:sp}

We describe how to infer the form of frequency dependence in an
exchangeable-allele Wright-Fisher model, from time-series data containing the
counts of alternative types over a sequence of generations.  We assume that the
fitness of any type at a frequency $p$ follows some function $w(p)$, where the
fitness $w$ is related to the corresponding selection coefficient $s$ via $w =
e^s$. The Wright-Fisher process allows us to write the transition probabilities
for exchangeable alleles of fitness $w(p)$ explicitly. Our goal is to infer the
fitness function $w(p)$ from the time series of observed transitions.

If we let $X_{i,t}$ be a vector random variable of the count of each type $i$
at time $t$ and let the index $i \in 1,...,k$ run over existing types
(temporarily disregarding mutation), the Wright-Fisher process evolves as the
stochastic process 
\begin{equation}
\label{eq:wfproc}
\bm{X}_{t + 1}|\bm{X}_t \sim
\operatorname{Multinom}(N_{t + 1}, 
[\pi_1,\pi_2,...,\pi_k]),\quad \pi_i =
{w(X_{i,t}/N_t)X_{i,t} \over \sum_{i=1}^k
w(X_{i,t}/N_t)X_{i,t}}.
\end{equation}
That is, $\bm{X}_{t + 1}$ given $\bm{X}_t$ is a multinomial sample of $N_{t +
1}$ total individuals, each having the probability $\pi_i$ to be of type $i$.
We note that $\pi_i$ here depends on $\bm{X}_t$ only, since $N_t$ is the total
population of types $1,...,k$ at at time $t$, $N_t = \sum_{i = 1}^k X_{i,t}$.
The fitness $w(p)$ is proportional to the expected per-capita reproductive
output of a type currently at frequency $p$ in the population prior to
enforcing any carrying capacity. That is, $w(p)$ is proportional to the
intrinsic growth rate and is hence a relative fitness, relative to an arbitrary
(unknown) standard: the multinomial probabilities in Eq.~\ref{eq:wfproc} are
unchanged if we multiply the fitness function $w(p)$ by a constant or,
equivalently, add a constant to the selection coefficients $s(p)$.

This exchangeable Wright-Fisher process thus specifies the probability of
observed counts $\bm{x}_0,...,\bm{x}_T$ over $T$ successive transitions between
generations as the product of conditional probabilities
\begin{equation*}
\Pr(\bm{X}_T = \bm{x}_T|\bm{X}_{T
- 1} = \bm{x}_{T - 1})\Pr(\bm{X}_{T - 1} = \bm{x}_{T - 1}|\bm{X}_{T - 2} = \bm{x}_{T -
2})\cdots\Pr(\bm{X}_1 = \bm{x}_1|\bm{X}_0 = \bm{x}_0).
\end{equation*}
Each conditional probability follows the multinomial distribution function,
$f(x, \bm{\pi})$
\begin{equation}
\label{eq:multilhd}
\Pr(\bm{X}_t=\bm{x}_t|\bm{X}_{t - 1} = \bm{x}_{t-1}) = f(\bm{x}_t,
\bm{\pi}(\bm{x}_{t-1})) = { n_{t}! \over
x_{1,t}!\cdots x_{k,t}!}\pi_1(\bm{x}_{t-1})^{x_{1,t}}\cdots
\pi_k(\bm{x}_{t-1})^{x_{k,t}},
\end{equation}
where $n_t = \sum_{i = 1}^k x_{i,t}$ is the total population size in generation
$t$. We write $\bm{\pi}(\bm{x}_{t-1})$ to emphasize that $\pi_i$ depends on
counts $\bm{x}$ (Eq.~\ref{eq:wfproc}). Viewed as a function of the observed
counts $\bm{x}$, $f$ is the distribution function, whereas as a function of
parameters, $f$ is the likelihood function.  Thus, given a realization of the
process $\bm{x}_t$, the likelihood is $\prod_{t=1}^{T} f(\bm{x}_t,
\bm{\pi}(\bm{x}_{t - 1}))$. When $w(p) = w(p|\bm\theta)$ can be parameterized
by a vector of smoothly-varying parameters $\bm\theta$, we can infer the
parameters using the method of maximum likelihood, up to an arbitrary constant
factor.

Finding the optimal parameter set, in particular when $w(p|\bm\theta)$ may have
many parameters, may be challenging. When the log-likelihood is concave, good
optimization algorithms are available. The likelihood of a sample path is the
product of the conditional likelihoods, and therefore the sample path log
likelihood has the same concavity the conditional likelihood. Taking the log of
Eq.~\ref{eq:multilhd} gives the log likelihood function
\begin{equation}
\mathcal{L}(\bm{x}|\bm\theta) = \ln n_t! - \sum_{i=1}^k \ln
x_{i,t}! + \sum_{i=1}^k
x_{i,t}\ln \pi_i(\bm{x}_{t-1}, \bm\theta),
\label{eq:multill}
\end{equation}
where we now write $\pi_i(\bm{x}_{t - 1},\bm\theta)$ to emphasize that by
Eq.~\ref{eq:wfproc}, $\pi_i$ depends on both $\bm{x}$ and on the parameters
$\bm\theta$ that determine the function $w(p)$.  To compute the derivative of
the log likelihood, we first write Eq.~\ref{eq:multill} in terms of $w$ and
bundle terms that are not a function of the parameters into a constant $c$. As
conditional likelihoods always involve a transition from one step to the next,
we use a prime ($'$) to denote the succeeding timestep, so that $\bm{x}_t$ and
$\bm{x}_{t - 1}$ are written $\bm{x}'$ and $\bm{x}$ respectively, and likewise
$n' = n_{t} = \sum_{i=1}^k x'_i$ and $n = n_{t -1} = \sum_{i=1}^k x_i$:
\begin{align}
\mathcal{L}(\bm{x}'|\bm\theta)  &= \ln n'! 
- \sum_{i=1}^k \ln x_i'! 
+ \sum_{i=1}^k x_i'\ln{x_iw(x_i/n|\bm\theta)\over\sum_jx_jw(x_j/n|\bm\theta)}
\nonumber \\
&= c + \sum_{i=1}^k 
  x_i'\ln{x_iw(x_i/n|\bm\theta)\over\sum_jx_jw(x_j/n|\bm\theta)} \nonumber \\
&= c + \sum_{i=1}^k x_i'\ln(x_iw(x_i/n|\bm\theta)) -
\sum_{i=1}^k x_i' \ln\left(
\sum_{j=1}^k x_jw(x_j/n|\bm\theta)\right)\nonumber \\
&= c + \sum_{i=1}^k x_i'\ln w(x_i/n|\bm\theta)
- n'\ln\left(\sum_jx_jw(x_j/n|\bm\theta)\right).
\label{eq:lumpedll}
\end{align}
Setting the derivative of the log likelihood with respect to $\bm\theta$ to
zero gives local extrema of the likelihood. The derivative is
\begin{align*}
&{d\over d\bm\theta}\left(
\sum_i x_i'\ln w(x_i/n,\bm\theta)
- n'\ln\left(\sum_ix_iw(x_i/n,\bm\theta)\right)  
\right)\\
&= \sum_i x_i' {{d \over d\bm\theta} w(x_i/n| \bm\theta) \over
w(x_i/n| \bm\theta)}
-n' {\sum_i x_i {d \over d\bm\theta} w(x_i/n| \bm\theta)
\over \sum_i x_i w(x_i/n| \bm\theta)} 
\end{align*}
and hence extrema exist when
\begin{equation}
\label{eq:ml}
 \sum_i x_i' {{d \over d\bm\theta} w(x_i/n|\bm\theta) \over
 w(x_i/n| \bm\theta)} =
n' {\sum_i x_i {d \over d\bm\theta} w(x_i/n|\bm\theta)
\over \sum_i x_i w(x_i/n|\bm\theta)}.
\end{equation}

\begin{figure}[t]
\noindent
\centering
\includegraphics[width=0.4\linewidth]{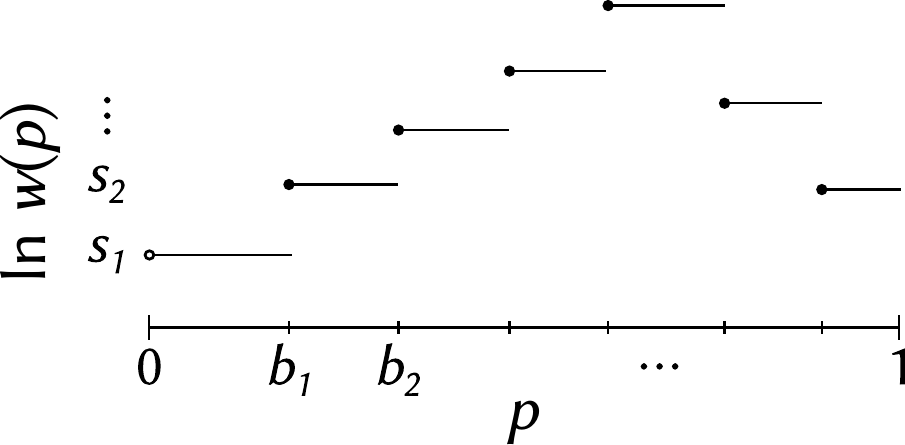}
\caption[Parameterization of $w(p)$]{ We let $w(p)$ be a piecewise constant
function with discontinuities at $b_d$, and selection coefficient $s_d$ within
each bin $d$. We let $\mathcal{B}(p)$ indicate the index $d$ such that $p \in
(l_d, u_d] = (b_{d - 1}, b_d]$.}
\label{fig:pwconstx}
\end{figure}

We parameterize $w(p)$ by assuming that $w$ is a piecewise-constant function of
$p$, where the parameters $\bm\theta = \{s_d\}$, $d \in 1,...,D$ are constant
selection coefficients $\{s_d\}$ associated with a set of frequency intervals
$\{(l_d,u_d]\}$ that are mutually-exclusive and exhaustive over the domain
$(0,1]$ of $w(p)$ (Fig.~\ref{fig:pwconstx}). Any given parameterization
specifies $D$ frequency bins with $D-1$ boundaries in $(0,1)$ labeled
$b_1,...,b_{D-1}$, so that 
\begin{equation}
\label{eq:param}
w(p|\bm{s}) = e^{s_{\mathcal{B}(p)}} = \begin{cases}
e^{s_1} & p \in (0, b_1]   \\
e^{s_2} & p \in (b_1, b_2] \\
\vdots  & \vdots           \\
e^{s_D} & p \in (b_{D-1},1]
\end{cases}
\end{equation}
For notational simplicity we let $b_D = 1$ and $b_0 = 0$ so that $(b_0,b_D]$ is
the full domain of $w(p)$, and we use $\mathcal{B}(p)$ to indicate the index
$d$ such that $p \in (l_d, u_d] = (b_{d - 1}, b_d]$. We thereby infer the most
likely fitness of types within each $s_d$ frequency interval. 

This $w(p)$ is a discrete frequency-dependent fitness function which is a good
approximation to continuous functions given a sufficient number of bins. The
bin boundaries themselves are an arbitrary choice and cannot be inferred using
maximum likelihood. Using finer frequency intervals gives successively better
approximations until too little data is available within each interval to get
sufficiently precise parameter estimates. This piecewise constant approximation
to the true relationship is analogous to image reconstruction: the resolution
limit on frequency bins is analogous to the trade-off between spatial precision
and noise in images.

What remains is to specialize the maximum-likelihood condition
(Eq.~\ref{eq:ml}) to our parameterization of $w(p)$ (Eq.~\ref{eq:param}), to
introduce mutation, and to solve for the maximum likelihood parameters given an
observed time series.

We deal first with mutation. The frequency $0$ is not in the domain of $w(p)$.
Types that are present in the current generation but were absent---i.e.~at
frequency $0$---in the previous generation must have arrived through some
process such as innovation or immigration. The infinite-alleles Wright-Fisher
process introduces new mutants stochastically at a constant rate $\mu$ per
individual per generation.  Each mutant is assumed to be of a completely new
type and given a new type identity so as to begin at initial frequency $1/N$
where $N$ is the total population size. In data, however, types not seen in the
preceding generation may appear at any frequency. We call the total count of
all such novel individuals appearing in a given generation $m_t$. We identify
these novel individuals with mutants in the infinite-alleles Wright-Fisher
process, and we show how to produce a maximum-likelihood estimate $\hat{\mu}$
of $\mu$. 

We call these types ``mutants'', but this is only a shorthand: we interpret
$\mu$ to be the rate of individuals with novel type appearing at whatever
initial frequency by whatever process (mutation, innovation, immigration,
proliferation within the data collection interval, etc.), assumed to be
independent of the current composition of the population. It is tempting in the
context of this study to identify $\hat{\mu}$ with an estimate of some rate of
cultural innovation, but our $\mu$ conflates many processes generating new
types in a population---including simply recalling names from a bygone era.

We modify the conditional likelihood expression to accommodate mutation,
incorporating $m_t$ observed new mutants and the parameter $\mu$. Again
focusing on the transition from generation $t-1$ to $t$, we use the prime
notation ($'$) to denote the later generation, but in so doing we take on a
change in semantics: absent mutation, we denoted total population size at time
$t$ as $n_t = n' = \sum_{i=1}^k x_i'$, where indices $1,...,k$ enumerate the
types present at time $t-1$. When mutations occur, however, $n_t$ cannot
simultaneously represent both the total population size at time $t$ and also
the sum of preexisting types, because the latter may exceed the former.  We
choose to let $n_t$ denote the total population size of all individuals, mutant
or otherwise, and retain the definition $n' = \sum_{i=1}^k x_i'$, so that $n'$
counts those individuals at time $t$ that arose from types present in the
previous generation. We define $m' = m_t$, and hence $n_t = n' + m'$ is the
total population size of all individuals at time $t$, including both
preexisting types and new mutants.  The new multinomial conditional log
likelihood for this transition between generations, analogous to
Eq.~\ref{eq:multill}, is given by
\begin{equation}
\label{multi-with-mut}
\mathcal{L}(\bm{x}'|\bm{s},\mu,\bm{x}) 
  = \ln (n' + m')! - \ln m'! - \sum_{i=1}^k \ln x'_i!
  + m'\ln \mu + \sum_{i = 1}^k x_i' \ln{(1 - \mu)x_iw(x_i/n) \over
  \sum_{j=1}^kx_jw(x_j/n)}
\end{equation}
The equivalence to Eq.~\ref{eq:multill} can be seen by imagining some $x_0' =
m'$ and a corresponding $\pi_0 = \mu$, rewriting the sums from $i=0$ to $k$,
and multiplying the $\pi_{i\ne0}$ by $(1-\mu)$ to retain unity in the sum of
probabilities. After rearranging terms and simplifying,
\begin{align}
\label{eq:fdlhd}
\mathcal{L}(\bm{x}'|\bm{s},\mu,\bm{x}) 
  &= \ln (n' + m')! - \ln m'! - \sum_{i=1}^k \ln x'_i!
\nonumber \\
  &+ m'\ln\mu + n'\ln(1-\mu)\\
  &+ \sum_{i = 1}^k x_i' \ln(x_iw(x_i/n|\bm\theta))
  - n'\ln\left(\sum_{j=1}^kx_jw(x_j/n|\bm\theta)\right). \nonumber 
\end{align}
Considering only terms involving $\bm{s}$, this log likelihood is equal to
Eq.~\ref{eq:lumpedll}: The additional terms involving $m'$ and $\mu$ do not
affect derivatives with respect to the parameters $\bm{s}$, and hence
Eq.~\ref{eq:ml} applies with respect to $\bm{s}$ regardless of the presence or
absence of mutation. The full set of parameters $\bm\theta$ however includes
the selection coefficients for each frequency bin as well as the mutation
rate, which we write $\bm\theta=(\mu,s_1, s_2, \ldots, s_D)$.

We maximize the likelihood by setting the derivatives with respect to all
parameters equal to zero. This results in a system of equations. The equation
for $\mu$ does not involve any other parameters and thus $\mu$ can be estimated
separately.
\begin{equation}
{\partial \over \partial \mu} \mathcal{L}
= {m'\over\mu} - {n'\over1 - \mu} = 0,
\label{eq:mlmu}
\end{equation}
This implies that the maximum-likelihood estimate $\hat\mu = m'/(n' + m')$ is
simply the observed fraction of mutants. Over the time series then,
${\partial \over \partial \mu}
\sum_{t=1}^T\mathcal{L}_t = 0$ implies
\begin{equation}
\hat\mu = {\sum_{t = 1}^Tm_t\over\sum_{t=1}^Tn_t}.
\label{eq:mlemu}
\end{equation}
The best estimate of mutation rate per individual per generation is simply the
observed fraction of mutants over the entire time series.

The condition for the maximum-likelihood parameter $\{\hat{s}_d\}$, then, is
only the set of equations ${\partial \over \partial s_d} \mathcal{L} = 0$ for
all $d \in \{1,...,D\}$.  Thankfully, the derivatives $dw/d\bm\theta$ that
appear in Eq.~\ref{eq:ml} are simply:
\begin{equation}
{\partial \over \partial s_d} w(p) =
\begin{cases}
e^{s_d} & \mathcal{B}(p)=d \\
0 & \mbox{otherwise}
\end{cases}
\end{equation}
Hence we can rewrite Eq.~\ref{eq:ml} using Eq.~\ref{eq:param},
$w(x_i/n|\bm\theta) = e^{s_{\mathcal{B}(x_i/n)}}$, and include terms $e^{s_d}$
for ${\partial \over \partial s_d}w(x_i/n|\bm\theta)$ whenever $x_i/n$ is in
bin $d$.  The factors $({\partial \over \partial s_d}w(x_i/n))/w(x_i/n)$ in the
left side of Eq.~\ref{eq:ml} are 1 when $\mathcal{B}(x_i/n) = d$ and 0
otherwise: They serves only to indicate for what $i$ to count $x'_i$ in the
sum.  Writing the maximum likelihood equations then mostly involves notational
bookkeeping about which terms to include in which sums.  Over a single
transition between subsequent generations, this amounts to the system of
simultaneous equations
\begin{equation} 
\sum_{i:\mathcal{B}(x_i/n) = d} \mkern-25mu x_i' =
\,\, n' \mkern-25mu \sum_{i:\mathcal{B}(x_i/n) = d} 
{x_i e^{s_d} \over \sum_{j=1}^k x_je^{s_{\mathcal{B}(x_j/n)}}}
\quad d \in \{1,...,D\}.
\label{eq:1genfdml}
\end{equation}
Because the log likelihood of the full time series is the sum of the conditional
log likelihoods, and because the derivative is a linear operator, the maximum
likelihood equations for the full time series are simply the sum of the
Eqs.~\ref{eq:1genfdml} over all transitions, namely
\begin{equation}
\sum_{t=1}^T
\left(\sum_{i:\mathcal{B}(x_{i,t-1}/n_{t-1})=d}
\mkern-25mu x_{i,t} \right)
= \sum_{t=1}^T
 \left[
(n_t - m_t)\left({
\sum_{i:\mathcal{B}(x_{i,t-1}/n_{t-1})=d} e^{s_d}x_{i,t-1}
\over
\sum_{j=1}^k
e^{s_{\mathcal{B}(x_{j,t-1}/n_{t-1})}}x_{j,t-1}}\right)\right]
\label{eq:fdml}
\end{equation}
for each $d \in {1,...,D}$.  This statement for the maximum-likelihood estimator
$\hat{\bm s}$, which satisfies these equations, has the interpretation of
``observed equals expected'': the left hand side is the observed counts
emanating from the $d$th frequency bin and the right side is the expected count
given the population configuration at time $t-1$ and number of mutations that
occurred between $t-1$ and $t$. That is, setting the expected data equal to the
observed data maximizes the likelihood.

The likelihood expressions already reveal features of the relationships between
the parameters $\mu$, $\bm{s}$ and $N$.  First, the mutation rate is
independent of the parameters $\bm{s}$. This is useful since the mechanisms of
mutation are often unknown or ill-defined in cultural contexts.  Second, the
derivatives of likelihood do not depend on $N$ except where $N$ normalizes
counts to frequencies.  Thus, the inference depends on the frequencies of
types, but not the total population size: The variance in frequency increments
are ignored.  This indicates that estimates of $s_d$ and estimates of effective
population size (by whatever means) are also independent. Thus, changing
timescales or population sizes (while suitably transforming units) has no
effect on the inference of $s_d$ in the diffusion limit assuming time intervals
are short.  Alternatively stated: if each increment of the data were replaced
by the same increment followed by $n$ generations of neutral Wright-Fisher
evolution, the expected value of the estimators of $s_d$ are unchanged.

For a single transition between subsequent generations, the system of equations
for the maximum-likelihood parameters (Eq.~\ref{eq:1genfdml}) is linear in
$e^{s_1}, \ldots, e^{s_D}$.  Nonetheless, the system of equations for the full
time series (Eq.~\ref{eq:fdml}) is difficult to solve, numerically or
otherwise, primarily because all $s_j$ appear in multiple different
denominators on the right hand side, which makes the equation for each
$\hat{s}_d$ non-linearly dependent on all other $\hat{s}_j$ in the general
case.

To the system of equations (Eq.~\ref{eq:fdml}) we use an MM or ``Minorize and
Maximize'' algorithmic strategy \citep{lange2000optimization}. The strategy is
to replace the likelihood $\mathcal{L}(\bm{s})$ of the parameter vector
$\bm{s}$ with a family of surrogate minorant functions $g(\bm{s}|\bm{s}^{(i)})$
where $\bm{s}^{(i)}$ is the $i$th iterate designed to approximate the maximum
likelihood estimate $\hat{\bm{s}}$ as $i$ becomes large. The minorant needs two
properties: $g(\bm{s}^{(i)}|\bm{s}^{(i)}) = \mathcal{L}(\bm{s}^{(i)})$ and
$g(\bm{s}|\bm{s}^{(i)}) \le \mathcal{L}(\bm{s})$. The minorant is guaranteed to
be at most $\mathcal{L}$ everywhere and to match $\mathcal{L}$ and its
derivatives with respect to $s_d$ at the point $\bm{s}^{(i)}$.  The consequence
is that either $\bm{s}^{(i)}$ is the optimum of $\mathcal{L}$, in which case
the derivatives of $g$ and $\mathcal{L}$ at $\bm{s}^{(i)}$ are zero, or else,
providing that $g$ and $\mathcal{L}$ are smooth, the derivative of $g$ is
non-zero and the optimum of $g$ lies somewhere between
$g(\bm{s}^{(i)}|\bm{s}^{(i)})$ and the optimum of $\mathcal{L}$. If we chose a
$g$ that is itself easy to optimize, we can monotonically approach the optimum
of $\mathcal{L}$ by finding the optimum $\bm{s}$ for each minorant
$g(\bm{s}|\bm{s}^{(i)})$ and choosing each successive optimum to be
$\bm{s}^{(i+1)}$. Each iteration finds a parameter set $\bm{s}^{(i)}$ that
closes the gap between $\mathcal{L}(\bm{s}^{(i)})$ and the true optimum of
$\mathcal{L}(\hat{\bm{s}})$.

The strategy requires an appropriate minorant. Many minorants are possible,
such as a quadratic approximation\cite{lange2010applied}. Examining the
likelihood function Eq.~\ref{eq:lumpedll}, we notice that the very last term,
\begin{equation*}
- n'\ln\left(\sum_ix_iw(x_i/n)\right)
= - n'\ln\left(\sum_ix_ie^{s_{\mathcal{B}(x_i/n)}}\right)
\end{equation*}
involves a sum inside the log, which gives rise to the troublesome denominator.
This term is a convex function of $w(p|\bm{s})$ and hence it is minorized by
its linear-order Taylor expansion in $w$ around $w(p|\bm{s}^{(i)})$:
\begin{equation}
-n'\ln\left(\sum_j{x_j} e^{s_{\mathcal{B}(x_j/n)}^{(i)}}\right)
-n'\sum_{d=1}^D\left(\left(e^{s_{d}} - e^{s_{d}^{(i)}}\right)
    {\partial\over\partial e^{s_{d}}}\left[
      \ln\left(\sum_{j=1}^k x_j
        e^{s_{\mathcal{B}(x_j/n)}}
      \right)
    \right]_{\bm{s}=\bm{s}^{(i)}}
  \right)
\label{eq:minorant}
\end{equation}
Substituting this linear minorant for the convex term in Eq.~\ref{eq:lumpedll}
produces a surrogate likelihood function $g(s|s^{(i)})$. We bundle terms that
are not dependent on $\bm{s}$ into a constant $c$, which includes those terms
that involve only $\bm{s}^{(i)}$ or $\mu$, $n$, et cetera:
\begin{align}
g(\bm{s}|\bm{s}^{(i)}) &= c + \sum_{d=1}^D \left [
  \sum_{j:\mathcal{B}(x_j/n)=d} \mkern-25mu x_j' s_{d}
  - n' e^{s_{d}}
    \left({
      \sum_{j:\mathcal{B}(x_j/n)=d}x_j
      \over
      \sum_{j=1}^k x_j e^{s^{(i)}_{\mathcal{B}(x_j/n)}}}
    \right)
	 \right ]
\label{eq:lumpedminorant}
\end{align}
The optimum $\bm{s}$, obtained by setting derivatives with respect to $s_d$ to
zero, is then given by the system of equations analogous to
Eq.~\ref{eq:1genfdml}:
\begin{equation}
\mkern-25mu \sum_{j:\mathcal{B}(x_{j}/n)=d} \mkern-25mu x_j' =
n'e^{s_d}
\left({
  \sum_{j:\mathcal{B}(x_j/n)=d}x_j
  \over
  \sum_{j=1}^k x_j e^{s^{(i)}_{\mathcal{B}(x_j/n)}}}
\right), \quad d \in \{1,...,D\}.
\label{eq:1genfdmm}
\end{equation}
Likewise, because a sum of conditional log likelihoods is minorized by a sum of
their respective minorants, Eq.~\ref{eq:fdml} has its analog, for $d \in
\{1,...,D\}$,
\begin{equation}
\sum_{t=1}^T
\left(\sum_{j:\mathcal{B}(x_{j,t-1}/n_{t-1})=d}
\mkern-25mu x_{j,t} \right)
=  e^{s_d} \sum_{t=1}^T
 \left[
(n_t - m_t)\left(
\sum_{j:\mathcal{B}(x_{j,t-1}/n_{t-1})=d}x_{j,t-1}
\over
\sum_{j=1}^k
x_{j,t-1}e^{s^{(i)}_{\mathcal{B}(x_{j,t-1}/n_{t-1})}}\right)\right].
\label{eq:fdmm}
\end{equation}
In fact, the only difference between the systems Eqs.~\ref{eq:fdml} and
\ref{eq:fdmm} is that the $s_j$ in the denominator of the right side have been
replaced by $s^{(i)}_j$. This difference is crucial, however, as the solution
for $\{s_d\}$ in the system of equations Eq.~\ref{eq:fdmm} is now trivial to
solve, which produces the next iterate:
\begin{align}
s_d^{(i+1)} &=
\ln \left[ \sum_{t=1}^T
\left(\sum_{j:\mathcal{B}(x_{j,t-1}/n_{t-1})=d}
\mkern-25mu x_{j,t} \right)\right] \nonumber \\
&-\ln \left(\sum_{t=1}^T
 \left[
(n_t - m_t)\left(
\sum_{j:\mathcal{B}(x_{j,t-1}/n_{t-1})=d}x_{j,t-1}
\over
\sum_{j=1}^k
x_{j,t-1}e^{s^{(i)}_{\mathcal{B}(x_{j,t-1}/n_{t-1})}}\right)\right]
\right).
\label{eq:mmiterates}
\end{align}
Here the right side of the equation involves only the previous iterate $\bm{s}^{(i)}$.

We initialize our MM algorithm by setting $s^{(0)}_d = 0$ for $d \in
\{1,...,D\}$ and we iterate until no further change in $\bm{s}^{(i)}$ is
achieved between iterations. That is, the derivative of the minorant at
$\bm{s}^{(i)}$ has reached approximately zero. At this point, the minorant and
$\mathcal{L}$ are both maximized.

As noted previously, the likelihood of the data are unchanged by adding a
constant to the selection coefficient $s_d$ in each frequency bin $d$. We
enforce the condition $\sum_{d=1}^D s_d=0$ to achieve a unique maximum
likelihood parameter set. We enforce the constraint by adding the appropriate
constant at each iteration. (Absent any enforcement of the constraint, the MM
iterates settle on an arbitrary value of $\hat{\bm{s}}$ that indeed satisfies
Eq.~\ref{eq:fdml}, but the particular value $\hat{\bm{s}}$ depends on the
initial condition $\bm{s}^{(0)}$.)

\subsection{Replacement fitness}

\label{sec:rep}

In Figure~\ref{fig:names} and elsewhere, we compute the replacement fitness,
$\bar{w}$.  The replacement fitness is a property of a time series, and it is
the fitness, which, if a type with that (constant, frequency-independent)
fitness were introduced at the beginning of the time series, its expected
frequency at the end of the time series would would equal its initial
frequency.

The population average fitness over time for generations $t \in \{1, ..., T\}$
and counts $x_{i,t}$ of types $i \in \{1, ..., k_t\}$ at generation $t$ is then
\begin{equation}
\label{eq:rep}
\bar{s} = \left({1 \over T}\right)\sum_{t=1}^T \log \left(\mu + \sum_{i = 1}^{k_t} 
(x_{i,t}/n_{t}) e^{s_{\mathcal{B}(x_{i,t}/n_{t})}}
\right).
\end{equation}

\subsection{Average frequency-dependent selection}

\label{sec:avg}

The average frequency-dependent selection can be computed from a realization of
an evolutionary process when the fitness $w_{i,t}$ of each type $i$ is known,
as in a simulation. Choosing frequency ranges as in Fig.~\ref{fig:pwconstx}, the
mean absolute fitness of types within a range $d$ is their expected per capita
total reproductive output over time,
\begin{equation}
\bar{w}_d = {\sum_t^T\sum_{i:\mathcal{B}(x_i/n)=d}w_{i,t}x_{i,t} \over
\sum_t^T\sum_{i:\mathcal{B}(x_i/n=d}x_{i,t}}
\end{equation}
The average selection within bin $d$, $\bar{s}_d$ is simply its log. In order
to set the zero point of selection relative to the replacement fitness, we also
subtract the replacement fitness, hence,
\begin{equation}
\bar{s}_d = \ln \bar{w}_d - \bar{s}
\end{equation}

\subsection{Wildtype fitness}

\label{sec:wildtype}

Sometimes we do not know exact counts of every type in the population or we
want to consider frequency-dependent fitness only within a subpopulation such
as biblical or non-biblical names. In these cases we make use of a wildtype
subpopulation with associated average fitness $s_w$, and treat all types
residing in that subpopulation (if they are known) as residing in their own bin
regardless of their frequency.  When counts are censored, for example, we do
not know exact counts for some types, but we may still know or estimate the
total number of censored individuals.  Since the sum of types in a given bin in
Eqs.~\ref{eq:fdml} is the same whether these types are labeled individually or
all belong to an aggregate type, we rewrite Eqs.~\ref{eq:fdml} in terms of
aggregate sums of wildtype individuals $w_t$ and $w'_t$ as follows for
frequency range $d$,
\begin{equation}
\sum_{t=1}^T
\left(\sum_{i:\mathcal{B}(x_{i,t-1}/n_{t-1})=d}
\mkern-25mu x_{i,t} \right)
= \sum_{t=1}^T
 \left[
(n_t - m_t)\left({
\sum_{i:\mathcal{B}(x_{i,t-1}/n_{t-1})=d} e^{s_d}x_{i,t-1}
\over
w_{t-1}e^{s_w} + \sum_{j=1}^k
e^{s_{\mathcal{B}(x_{j,t-1}/n_{t-1})}}x_{j,t-1}}\right)\right]
\end{equation}
taken simultaneously with the following equation for the wildtype fitness
$s_w$,
\begin{equation}
\sum_{t=1}^T
\left(w'_{t-1}\right)
= \sum_{t=1}^T
 \left[
(n_t - m_t)\left({
e^{s_w}w_{t-1}
\over
w_{t-1}e^{s_w} + \sum_{j=1}^k
e^{s_{\mathcal{B}(x_{j,t-1}/n_{t-1})}}x_{j,t-1}}\right)\right].
\end{equation}
The wildtype aggregation $w_{t-1}$ is the total number of wildtype individuals
at generation $t-1$, and $w'_{t-1}$ is the total number of generation-$t$
progeny of wildtype individuals in generation $t-1$.  One might expect
$w'_{t-1}$ to equal $w_t$, yet the two accountings may differ. As an example
suppose we let the wildtype represent types at frequency below $f$: if a type
transitions from below $f$ at generation $t-1$ to above $f$ at generation $t$,
it contributes to $w_{t-1}$ and $w'_{t-1}$ but not to $w_t$. The aggregates
support an exact analogy between the wildtype counts $w_t$, wildtype progeny
$w'_t$ and wildtype fitness $s_w$ and the counts,
$\sum_{i:\mathcal{B}(x_{i,t-1}/n_{t-1})=d} x_{i,t-1}$, progeny
$\sum_{i:\mathcal{B}(x_{i,t-1}/n_{t-1})=d} x_{i,t}$, and fitness $s_d$ of
frequency-range $d$. Thus the wildtype aggregate acts as a
frequency-independent bin, and we estimate its fitness in the same way.

\subsection{Subpopulations}

\label{sec:subpop}

\begin{figure}[t] 
\noindent
\centering
\includegraphics[width=120mm]{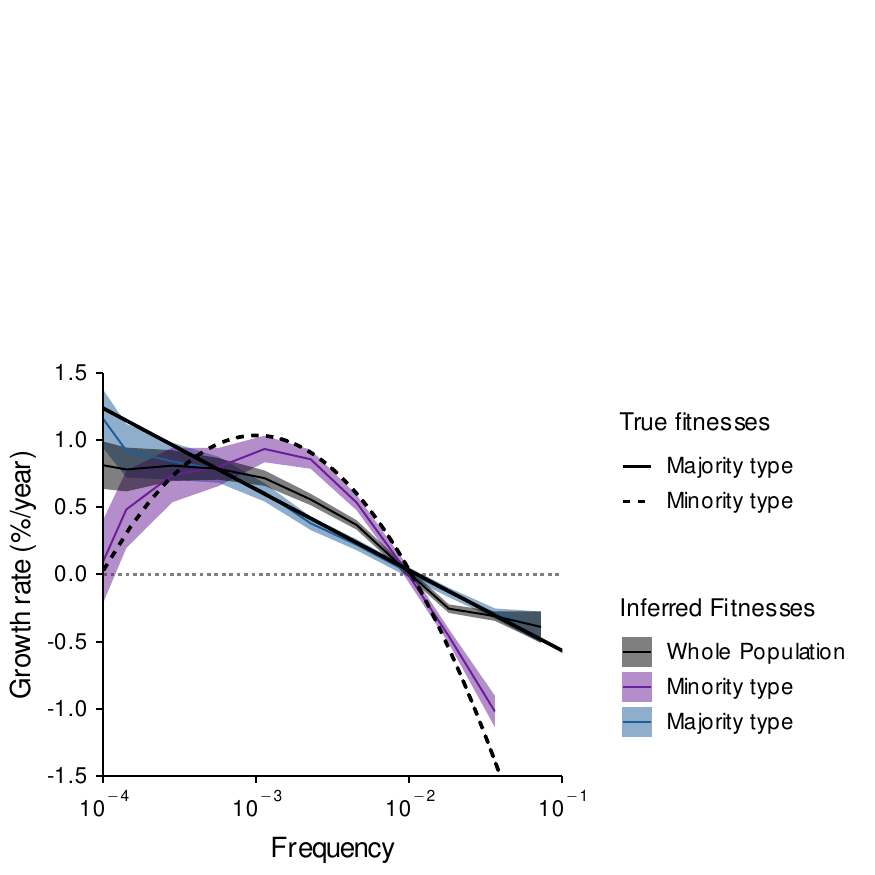}
\caption[]{\textbf{Inferring subpopulation frequency-dependent fitness from
simulation.} We recover subpopulation frequency-dependent fitness from a
simulated population composed of two subpopulations: a minority type (one third
of mutants) with quadratic frequency-dependent fitness $w(p) = 1 -
0.01(\log_{10}p + 4)(\log_{10}p + 2)$, and a majority type (two thirds of
mutants) with linear frequency-dependent fitness $w(p) = 1.012 -
0.006(\log_{10}p + 4)$. Both forms have a zero-growth intercept at $10^{-2}$.
We simulate a Wright-Fisher process with fixed population size 20,000 and
mutation rate 0.0003 for 50,000 generations after 20,000 generations of burn-in
from a monomorphic initial population. Confidence bands are analytic 95\%
confidence intervals derived from Fisher information (\ref{sec:cis}).}
\label{fig:parlin}
\end{figure}

Wildtype aggregations (\ref{sec:wildtype}) also allow measurement of
frequency-dependent selection among subpopulations. By treating all individuals
who are not part of the focal subpopulation as part of the wildtype
aggregation, those individuals do not contribute to inferred
frequency-dependent selection, but do contribute to replacement fitness. Hence
separate estimates for multiple mutually-exclusive subpopulations will be
calibrated against the same replacement fitness as demonstrated in
Fig~\ref{fig:parlin}.

\subsection{Confidence intervals}

\label{sec:cis}

We use two confidence intervals (CIs) in this study: a CI derived from the
model and a CI based on a bootstrap. The model CI is an accurate theoretical
prediction of the variance on the estimator subject to the assumption that the
data is generated by the model. The bootstrap CI is 2.5\% to 97.5\% quantiles
of a non-parametric bootstrap \citep{efron1986bootstrap} that generates an
empirical distribution of the estimator by resampling the underlying data. A
bootstrap CI makes no assumption that the data are generated from the assumed
model, but does assume that a resampling scheme is available that generates
equally probable samples, such as multinomially resampling the observed
dataset. Either of these assumptions may fail in practice. Corroboration
between the two CIs, as we observe, supports the accuracy of the model and the
underlying assumptions. When the CIs disagree, the bootstrap CI is more
conservative, as it accounts for some failures of the model to describe the
data.

For the large samples used to infer dog breeds and baby names, we use model
confidence intervals on parameter inferences computed using the observed Fisher
information \citep{efron1978assessing}, a measure of the curvature of the
likelihood function at its maximum. With an identifiable parameter in the limit
of many independent observations, the log-likelihood surface is approximately
parabolic near its maximum. The observed information is the matrix of second
partial derivatives of the likelihood function (Hessian matrix), evaluated at
the most likely parameters, and its inverse is the variance-covariance matrix
of the estimators $\{\hat{s}_d\}$. We compute 95\% confidence intervals on each
parameter as $1.96\sigma_{\hat{s}_d}$, where $\sigma_{\hat{s}_d}$ is simply the
square root of the diagonal elements in the observed information matrix.  As
noted above, we enforce the relationship  $\sum_{d=1}^D \hat{s}_d=0$ among the
inferred selection coefficients. And so, in practice, the Hessian has
dimensions $(D-1) \times (D-1)$, corresponding to $s_1, \ldots s_{D-1}$. We
compute the corresponding variance in the estimator $\hat{s}_D$ using the
relationship $\text{var}(\hat{s}_D) = \sum_{d=1}^{D-1} \text{var}(\hat{s}_d) +
2 \sum_{0 < e < d < D} \text{covar}(\hat{s}_e,\hat{s}_d)$.

We verified the accuracy of our parametric confidence intervals by simulation,
using a parametric bootstrap.  In particular, we simulated the model using US
inferred parameters to generate samples equal in size to our data, and we
verified that the 95\% confidence intervals derived from observed Fisher
information include the true parameters of the simulations 95\% of the time
after 1,000 trials, within binomial error.

We compute bootstrap CIs by resampling transitions within the time series.
Bootstraps that accommodate the dependence relationships that may be present in
time series is an open-ended topic \citep{politis2003impact}, because generally
there is no resampling scheme that logically guarantees resampled data to be
equiprobable under the true generating process, in contrast to the bootstrap
argument for independent samples.  Nonetheless, in our case, the likelihood
calculation sums over observations and expectations of change from one timestep
to the next, and it is these transitions that are assumed to be independent
across time periods and approximately independent between types. Unaccounted
dependence between observations enters by either failures of the Markovian
assumption in the true generating process, or by effects of competition between
types induced by finite population size.  We resample in a way that
approximates independence and roughly preserves population size so that
resampled transitions have similar frequencies to the originals.

We resample a given timestep by choosing at each timestep transitions $x_i \to
x'_i$ from that timestep in the data uniformly with replacement.  We determine
the number of transitions to include at each timestep by iteratively including
another transition sample with probability $\max(1, p - c / e - c)$ where $c$
is population size of the types included thus far, $p$ is the observed
population size in the original data, and $e$ is the expected population size
increase by including one more transition sample.  This guarantees that the
expected resampled population size is equal to the observed population size,
but allows statistical variation in population size from generation to
generation. This resampling scheme is equivalent to a multinomial resampling of
independent observations which accounts for robustness of the estimator to
sampling noise, but also preserves the frequency interpretation of the original
count data.  The frequency associated with a given count is allowed to vary
around its expectation in the bootstrap, and thus the bootstrap also represents
robustness of the estimator to the inclusion or omission of observations due to
sampling of the data, since such sampling omissions would cause the frequencies
in observed data to fluctuate.  The bootstrap CIs are conservative estimates of
the true 95\% CI.

\subsection{Accounting for Censorship}

\label{sec:cen}

We might naively assume that time series are complete and uncensored. In the
case of baby names however, most datasets censor annual counts below a certain
threshold. In the United States (US), for example, the Social Security
Administration censors, each year, any name that occurred fewer than five times.
To be conservative with respect to biases due to censorship, we do not
calculate or report selection coefficients for frequencies so small that they
would correspond to censored counts or counts less than one for any generation
in a time series. This frequency is
\begin{equation*}
f_{\min} = \max_t f_{\min,t}
\end{equation*}
where $f_{\min,t}$ is the minimum frequency that would still be rounded to an
uncensored count at generation $t$, that is $f_{\min,t} = c - 0.5 / n_t$ where
$c$ is the minimum uncensored count (e.g. $c = 5$ for US or $c=1$ if there is
no censorship).

We do not ascribe counts at frequency below $f_{\min}$ to any frequency range,
ascribing them instead to the wildtype aggregation (See \ref{sec:wildtype}). We
count types at frequency below $f_{\min}$ among the wildtype---that is the sum
$w_{t-1}$ includes all counts $x_{i,t-1}$ of types $i$ existing in the data at
generation $t-1$ such that $x_{i,t-1}/n_{t-1} < f_{\min}$. Likewise $w'_{t-1}$
includes both progeny of those wildtype individuals as well as the total count
$m_t$---the sum of all types that did not appear in the data in generation
$t-1$---which may be either mutant/immigrant types or progeny of unobserved
types.  Censorship therefore conflates true ``mutants'' with progeny of types
at too low frequency to observe. This entails a new interpretation of mutation:
rather than consider types appearing for the first time in the dataset as
mutants arising from count zero, we simply consider them as types arising from
some frequency too low to be observed in the data. Thus the wildtype fitness
$s_w$ accounts for the relative competitiveness (growth rate) of mutants,
migrants or innovations as well as the progeny of types of unknown frequency or
frequency below $f_{\min}$. Hence no mutation term enters the replacement
fitness calculation (Eq.~\ref{eq:rep}) because the affect of mutation on
replacement fitness is subsumed into the contribution of the average fitness
$s_w$ of wildtype individuals.

Censorship also biases inferred selection coefficients for frequency ranges
above $f_{\min}$, because transitions from a recordable frequency $x >
f_{\min}$ to a lower frequency may not be representable in the data or may be
censored, whereas transitions from $x$ to a higher frequency are always
recorded. The consequence for the associated selection coefficient depends on
our interpretation of types that disappear from observation. If we
``pessimistically'' interpret the disappearance of a type from the data as a
transition from frequency $x$ to frequency 0, this negatively biases the
associated selection coefficient by substituting zero for its true frequency,
which could be anywhere between zero and the minimum observable frequency.
This pessimistic interpretation provides a lower bound on the inferred
selection coefficient associated with frequency $x$. The corresponding
``optimistic'' interpretation or upper bound imputes the disappearance of a
type as a transition to the minimum uncensored count---guaranteed to be greater
than the maximum unobservable frequency.

We combine these upper and lower bound interpretations for individual
transitions to produce ``optimistic'' and ``pessimistic'' interpretations of
the full frequency-dependent selection curve by interpreting all partially
observed transitions either optimistically or pessimistically. These
interpretations are not strict upper or lower bounds on the selection
coefficients per se, because transitions do not affect selection coefficients
superpositively: an upper-bound interpretation of a transition from frequency
$x$ may lower the selection coefficient for frequency $y \ne x$. However,
because the dependence between inferences at different frequencies is typically
weak, we treat the optimistic and pessimistic interpretations as approximate
upper and lower bounds on all selection coefficients.

To compute the optimistic and pessimistic interpretations of the censored
counts, we adjust Eq.~\ref{eq:fdml} to account for types at frequency below
$f_{\min}$ as wildtype and to impute pessimistic and optimistic type counts.
We use the following expression, which differs from Eq.~\ref{eq:fdml} in that
$m_t$ is subsumed into $w'_t$, $n_t$ has been replaced with $\hat{n}_t$, and
some $x_{i,t}$ have been replaced with $\hat{x}_{i,t}$ as described below:
\begin{equation*}
\sum_{t=1}^T
\left(\sum_{i:\mathcal{B}(x_{i,t-1}/\hat{n}_{t-1})=d}
\mkern-25mu \hat{x}_{i,t} \right)
= \sum_{t=1}^T
 \left[
\hat{n}_t\left({
\sum_{i:\mathcal{B}(x_{i,t-1}/\hat{n}_{t-1})=d} e^{s_d}x_{i,t-1}
\over
w_{t-1}e^{s_w} + \sum_{j=1}^k
e^{s_{\mathcal{B}(x_{j,t-1}/\hat{n}_{t-1})}}x_{j,t-1}}\right)\right]
\end{equation*}
taken simultaneously with the following equation for the wildtype fitness
$s_w$,
\begin{equation*}
\sum_{t=1}^T
\left(w'_{t-1}\right)
= \sum_{t=1}^T
 \left[
\hat{n}_t\left({
e^{s_w}w_{t-1}
\over
w_{t-1}e^{s_w} + \sum_{j=1}^k
e^{s_{\mathcal{B}(x_{j,t-1}/\hat{n}_{t-1})}}x_{j,t-1}}\right)\right].
\end{equation*}
The hatted quantities differ from what they replace as follows. The $\hat{n}_t$
denotes the imputed total population size including individuals that are
censored from the data, either by including known aggregate censored counts,
independent measurements of total population size, or by imputing the full,
uncensored population size.  The $\hat{x}_{i,t}$ is simply $x_{i,t}$ if it is
present in the dataset, otherwise it is the imputed count that depends whether
we compute the optimistic or pessimistic imputation. Under the pessimistic
imputation, $\hat{x}_{i,t} = x_{i,t} = 0$. Under the optimistic imputation,
$\hat{x}_{i,t} = c$ where $c$ is the minimum uncensored count possible in the
dataset at that generation. It must be the case for all $t$ that
\begin{equation*}
w'_{t-1} + \mkern-15mu
\underset{i:\mathcal{B}(x_{i,t-1}/\hat{n}_{t-1})=d
}{\mkern-55mu \sum_{d\in 1,...,k_t} \mkern-15mu \sum}
\mkern-50mu \hat{x}_{i,t} 
=
\hat{n}_t,
\end{equation*}
where $k_t$ is the number of types present in data at time $t$.  We use this
equation to define $w_{t-1}'$ in terms of $\hat{n}_t$ and the $\hat{x}_{i,t}$.
It follows that
\begin{equation*}
w'_{t-1} > \mkern-30mu
\sum_{i:\hat{x}_{i,t}/\hat{n}_{t} < f_{\min}}
\mkern-30mu \hat{x}_{i,t}
\end{equation*}
for types $i$ occurring in the data. The range of the sum, the types
$i:\hat{x}_{i,t}/\hat{n}_{t} < f_{\min}$, is the exact complement of types
assigned to any frequency bin among types present in the data at generation
$t$.  Furthermore,
\begin{equation*}
w_{t-1} + \sum_{j=1}^{k_{t-1}} x_{j,t-1} = \hat{n}_{t - 1},
\end{equation*}
and hence we use this as a definition of $w_{t-1}$ in terms of $\hat{n}_{t-1}$
and types that appear in the data at frequencies greater than $f_{\min}$. Hence
$w_{t-1}$ count both types that appear in the data at frequencies below
$f_{\min}$ and types that do not appear in the dataset at all.

The two interpretations of the missing data give high and low estimates for
selection coefficients on rare types. However, the fitness ascribed to rare
types affects replacement fitness to some extent, so the optimistic estimate
for rare types often leads to slightly lower estimates for common types through
increasing the replacement fitness. In practice we use the spread between the
estimates across the plotted region as a measurement of potential bias due to
censorship and conflation of mutants with offspring of low-frequency types.

\subsection{Sampling biases}

\label{sec:samp}

The real process of naming and the Wright-Fisher process involve similar
sampling noise, but a subtle discrepancy in the noise structure can causes bias
in estimates, particularly at low frequencies. In the Wright-Fisher process,
sampling noise (variance in frequencies from generation to generation) is
exactly the multinomial variance due to sampling with replacement from the
types present in the previous generation. The real process however samples
names not strictly from the previous generation, but from some larger,
unobserved population of names in an evolving cultural milieu. The discrepancy
sometimes causes measurable bias in estimated selection coefficients,
particularly frequencies corresponding to low absolute counts. For example, the
underlying Wright-Fisher process assumes exact lineal continuity from
generation to generation, such as would be the case with last names or genes
attributable to a particular parent.  First names, rather, are copied from many
sources, including bygone generations. Hence, some sequences that occur in the
data are impossible under the Wright-Fisher process, such as when a name
apparently goes extinct then reappears. Aside from being of statistical
interest, this very discrepancy has interesting philosophical and scientific
implications which we discuss at the end of this section.

We control for this bias by constructing time series of samples in which the
only change from generation to generation is sampling noise, then we infer
frequency-dependent selection from the time series of samples. First, we
construct a fixed, time-independent metapopulation composed of all counts of
all types summed over the duration of the dataset. Then we construct an
artificial time series based on repeated samples with replacement from the
metapopulation, matching population size to the data each generation.  The
average frequency of every type at every generation in the time series of
samples is just its frequency in the metapopulation, and therefore no type
changes frequency over time on average. Hence, in principle, there can be no
selection and no frequency dependent growth. We then infer frequency-dependent
selection in this time series of samples and take the magnitude of departure
from neutrality to estimate the magnitude of potential bias. We typically find
substantial bias at low frequencies where binomial sampling noise predominates.

Inferred frequency dependence in the time series of samples is possible because
of apparent growth or decline from generation to generation due to fluctuations
in sampled counts: If a type happens to be sampled below its frequency in the
metapopulation, it appears to grow on average in the next generation, and the
opposite occurs if it is sampled above its true frequency. Typically, these
generation-to-generation fluctuations cancel over time and the inference
converges to zero selection. However, in certain circumstances cancellation
does not occur. For example, when a type appears to go extinct and then
reappears, its decrease to extinction registers as negative selection and
whereas its subsequent increases in registers as mutation or immigration rather
than positive selection, causing drastic bias towards negative selection. A
qualitatively similar phenomenon occurs if any type fluctuates below or above a
bin boundary from generation to generation. The likelihood of these occurrences
diminishes as sampling variance decreases with higher counts.

The net effect of sampling bias depends on the total population size, the bin
boundaries, and the distribution of type frequencies. We do not attempt to
predict the bias analytically. Instead we simply infer frequency dependence in
the resampled time series to determine at what frequencies this bias exists.
Experimentally, the bias typically only exists at frequencies corresponding to
low counts, with the exact frequencies affected varying by dataset
(Fig~\ref{fig:samp}).

Finally, differences between cultural and genetic evolution spurred
debate\cite{lewens2015cultural,acerbi2015if} about how far a cultural Darwinian
metaphor can be carried.  Chief among the differences are starkly contrasting
mechanisms of inheritance\cite{mesoudi2011cultural}.  Extinctions of species
and genes are permanent, whereas languages with no speakers have been
resurrected, leading to complications in interpreting ``extinction'' of
cultural traits\cite{krauss2007classification}.  Whereas family names and
lineages may go extinct, a first name, word or language may persist in a
cultural repertoire long after its last recorded instance in any dataset. The
sampling bias we observe in names is instructive in exactly to what degree
these differences are relevant to population-level change. On the one hand, the
data manifestly violates the Wright-Fisher process because even a complete
birth cohort of names is still a sample from a larger cultural repertoire, and
absence of a name from one birth cohort does not imply absence in a subsequent
cohort. On the other hand, the differences are negligible above a certain
frequency and the dynamics of name frequencies can be meaningfully treated as a
Wright-Fisher diffusion. The differences confuse measurements of mutation and
extinction, but in measuring frequency-dependent selection, the discrepancies
between cultural and genetic models only matter near the boundary at 0
frequency, where even alternative population-genetic models often behave
differently\cite{durrett2008probability}.

\subsection{Inferring $N_e$}

Our primary concern thus far has been estimating fitness as a function of
frequency, which determines the mean change (advection) in the diffusion
equation.  The diffusion term is also of interest.  In the diffusion limit of
the Wright-Fisher process, the diffusion term is $(1/2)\partial^2/\partial x^2
(x(1-x)/N)\phi(x,t)$, where $N$ is the number of allele copies in the
Wright-Fisher process \citep{kimura1964diffusion}, and $\phi(x,t)$ is the
probability density at frequency $x$ at time $t$.  The coefficient $x(1-x)/N$
describes that variance in gene frequency accrues at a rate of one binomial
sampling per generation, and has the units of a transport coefficient:
frequency$^2/$time. The parameter $N$ represents the variance effective
population size, and it equals the census population size (or its geometric
mean if it fluctuates) when one round of random sampling (a single multinomial
draw) occurs at each time interval, $g$, i.e., one generation. For short time
intervals, variance accrues linearly in time in the diffusion equation, and so
if sampling is conducted at a different interval $g'$, $N'$ will be rescaled
proportionately such that $g/g' = N'/N$.  Thus, at sampling interval $g$ and
census population size $N$, if the inferred effective population size $N_e$
differs from $N$, then there is an effective generation time $g_e = gN/N_e$
which is the time for the diffusion to accrue the same variance as one
Wright-Fisher generation---one multinomial sample of the full census
population.

We infer $N_e$ by rescaling frequency increments in order to produce
homoscedastic updates under the inferred model parameters.  If $r_i$ are the
residuals of the data, $r_i = x_i - e_i$, where $x_i$ are the data and $e_i$
are the expectations under the model, then the rescaled increments are
\begin{equation}
Y_i = {r_i \over \sqrt{x_i(1-x_i)}}.
\end{equation}
The rescaled increments $Y_i$ have variance $1/N_e$ per generation, so $N_e =
1/S$ where $S$ is the sample variance of the $Y_i$.

We estimate the effective population size $N_e$ and a generation time for baby
names in each country in Table~\ref{tab:babstats}, where the generation time
represents the characteristic timescale on which complete mixing of names
occurs.

\section{Data Handling and Error Estimation}

\begin{table}[t]
\centering
{\sffamily\fontsize{7pt}{10pt}\selectfont
\begin{tabular}{l r r r r r r r}
\medskip\\
Name Data & Years used & Total indiv. & $\hat{N}_e$ &
$\hat{N}_{e,s=0}$ & Censorship & Cens. err. & Samp. err. \\
\hline
United States &
1936--2018 &
305,742,609 &
435,541 &
430,075 &
$<$5 births/year &
$\le$0.01 &
$\le$0.08 \\
France &
1946--2018 &
60,314,639 &
57,332 &
57,434 &
$<$20 or $<$4/year &
$\le$0.02 &
$\le$0.39 \\
Netherlands (1y) &
1946--2014 &
15,408,439 &
112,919 &
137,813 &
none &
--- &
$\le$0.39 \\
Netherlands (5y) &
1950--2010 &
14,351,523 &
147,453 &
145,038 &
none &
--- &
$\le$0.04 \\
Norway &
1946--2019 &
4,533,582 &
20,673 &
20,322 &
$<$4 births/year &
$\le$0.08 &
$\le$0.59*\\
\hline
\end{tabular}
*discounting anomalous control in highest frequency
bin.}
\label{tab:babstats}
\caption[Name data summary]{\textbf{Name dataset summary.} For each name
dataset used, the years, total number of individuals, inferred
variance-effective population sizes $\hat{N}_e$ and $\hat{N}_{e,s=0}$ assuming
$s(p)=0$ computed from model residuals, the level or type of censorship, the
maximum sampling bias within in plotted range (in \%/year), and the maximum
censorship bias within the plotted range (in \%/year).}
\end{table} 

\subsection{Names}

\label{sec:names}

For United States baby names, we obtained public data from the United States
Social Security Administration database of first names
(\url{https://www.ssa.gov/OACT/babynames/}) 1880-2018.  We omitted nine high
frequency names that clearly resulted from coding errors around 1985 such as
``Christop,M'' and ``Infant,F''. We omitted data up to and including 1935 as
non-representative of a complete population of individuals. Births in the
database that occur prior to 1935 represent cards issued to children and
adults, as opposed to the post-1935 practice of issuing cards at birth. Hence
the data prior to 1935 are incomplete and contain unrealistically small birth
cohorts. We consider as distinct names any spelling variants or the identical
spelling in persons of different sex. The dataset indicates birth sex as a
suffix on the name (either ``,M'' or ``,F''). We conducted the inference
considering each annual birth cohort to be a generation of the Wright-Fisher
process, and infer parameters for $\hat{s}(p)$ composed of logarithmically
increasing frequency ranges (bins) starting from 0.0001-0.0002 with bin
boundaries increasing by powers of two (0.0001, 0.0002, 0.0004, etc.) following
Hahn and Bentley\cite{hahn2003drift}.  We additionally infer but do not report
fitness parameters for names more rare than 0.0001 whose frequency may or may
not be known as described below.

The US data censors counts below 5. Hence, computing an accurate mutation rate
is impossible and some fraction of individuals are missing (censored) from the
dataset.  To account for unobserved individuals, we conduct the inference using
a frequency-independent wildtype fitness for censored, mutant and low-frequency
types (\ref{sec:wildtype}, \ref{sec:cen}). We guess the fraction of missing
individuals in the time series by imputing a \texttt{CENSORED} type which always
counts 5\% of the sum of the uncensored population each year.  We derived the
number 5\% from the French dataset, in which aggregate counts of censored
individuals are listed explicitly and the fraction of counts below 5 is known
to be 4.5\%. In the US dataset the true count of names at frequency below
$f_{\min}$ (censored and uncensored) far exceed the sum of those names that
appear uncensored in the data below $f_{\min}$.  As described in \ref{sec:cen},
we pool the \texttt{CENSORED} type together with uncensored counts less than
the maximum censored frequency $f_{\min}$=2.1e-06 into the wildtype
subpopulation. The inferred wildtype fitness explicitly conflates immigrant
types, mutant types, and progeny of types with censored counts into an average
growth rate of all unaccounted types. This wildtype fitness is necessary to
compute the replacement fitness or zero point of growth but otherwise has no
affect on reported results. The error in replacement fitness due to censorship
and conflation of mutants with progeny of unobserved types is measured in the
spread between optimistic and pessimistic selection coefficients at high
frequencies and reported below.

\begin{figure}[t]
\noindent
\centering
\includegraphics[width=120mm]{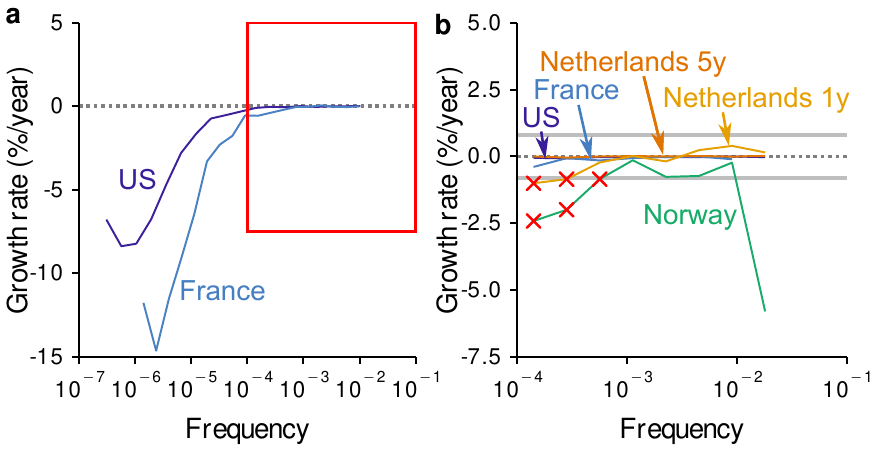}
\caption[Sampling bias controls by
country.]{\textbf{Sampling bias controls by country.} Inference from synthetic
time series composed of generations sampled with replacement from a static
metapopulation of name frequencies derived from the data. Inference with 20
bins (a) at all frequencies shows bias at low frequencies in the US and France.
The red box indicates the plot range of Fig~\ref{fig:names}.  Panel (b) shows
the control inference for the same bins and plot range as Fig~\ref{fig:names}.
Red Xs indicate bins removed from the results for exceeding the error threshold
(b, grey lines). We ignore the anomalous control in the highest bin in Norway:
It is an artifact that occurs because the highest-frequency name in the
metapopulation (Anne) is only slightly less than the bin boundary and
occasionally passes into the highest bin only to return, registering low
fitness.}
\label{fig:samp}
\end{figure}

We controlled for sampling bias by constructing time series of samples from the
time-independent distribution of name frequencies as described in
\ref{sec:samp}. The inference from this synthetic time series
(Fig.~\ref{fig:samp}) measures the magnitude of possible bias due to sampling
unaccounted in the Wright-Fisher process.

At frequencies greater than 0.0001 in the US inference, censorship bias affects
plotted frequencies by at most 0.06\%/year and sampling bias by at most
0.08\%/year although frequencies less than 0.0001 were affected by both
censorship bias (\ref{sec:cen}) and sampling bias (\ref{sec:samp},
Fig.~\ref{fig:samp}). Hence, we do not report inferred fitnesses for names
below frequency 0.0001.

We further acquired baby name data from France 1900-2018 (Insee Fichier des
prénoms\footnote{\url{https://www.insee.fr/fr/statistiques/fichier/2540004/nat2018_csv.zip}},
retrieved July 3, 2020) and Norway 1880-2019 (Statistics Norway Names
STATBANK\footnote{\url{https://www.ssb.no/statistikkbanken/selectvarval/Define.asp?SubjectCode=al&ProductId=al&MainTable=FornavnFodte&SubTable=1&PLanguage=1&Qid=0&nvl=True&mt=1&pm=&gruppe1=Hele&aggreg1=&VS1=NavnJenter02&CMSSubjectArea=befolkning&KortNavnWeb=navn&StatVariant=&TabStrip=Select&checked=true}},
retrieved Aug 18, 2020). Additionally we received complete, uncensored counts
from a name corpus\cite{bloothooft2015corpus} of the Netherlands 1946-2015
anonymized by Gerrit Bloothooft of Meertens Instituut and communicated by
email. In France data, we ignored inexhaustive data prior to 1946 as indicated
in the file metadata. France censors names that do not occur at least 20 times
since 1946 and names that occur less than 3 times in any given year, but lists
the totals of censored counts in years \texttt{XXXX} and types
\texttt{\_PRENOMS\_RARES}. We incorporated the censored counts into a wildtype
aggregation with counts below $f_{\min}$=3.4e-06, as with the US inference
(\ref{sec:cen}), distributing the counts of unknown year uniformly across the
dataset. In Norway data, we ignored records prior to 1946 because many
contained missing values. Counts less than 4 are censored in Norway, so we
again assumed 5\% of the data to be censored in each year and incorporated the
imputed censored counts into the wildtype aggregate of types below
$f_{\min}$=7.8e-05.

We excluded from plots any bins with less than 5 types or censorship or
sampling bias exceeding 0.8\%.  Censorship bias in France and Norway was
minimal except in the rarest frequencies in Norway, but sampling bias was
substantial in both Norway and the Netherlands due to their small sizes. Hence,
we first excluded any frequency bins for which the sampling bias control
exceeded 0.8\%/year (Fig.~\ref{fig:samp}) with the exception of the
highest-frequency bin in Norway. In the remaining (plotted) bins in all
countries, censorship bias (as measured by the difference between optimistic
pessimistic interpretations of missing data) was at most 0.08\%/year.

With the exception of Norway, the sampling bias control for displayed bins was
at most 0.4\%/year.  Norway was most affected by sampling error due to its
small size. We kept the highest bin in Norway despite its control exceeding
0.8\%/year because the control is anomalous: The most frequent name in Norway,
``Anne'', has a frequency 0.0127 over all time, just under the highest
frequency bin boundary of 0.0128. This name is the only name to occupy the
highest bin in the time series of samples, entering the bin 7 times always to
return the following timestep, registering negative growth. In the data by
contrast, 17 names populate the highest frequency bin for sustained time
periods, so we do not believe this bin is truly substantially affected by
sampling bias. The root mean square RMS bias in Norway with the exception of
the highest bin was 0.5\%/year whereas for all other plotted bins combined the
RMS bias was 0.14\%/year.  Finally, we observe from Fig.~\ref{fig:samp} that
sampling bias does not substantially affect the fitness of common types and by
extension the replacement fitness.

Lastly, there is error associated with sampling at finite time intervals. We
conduct inference using the Wright-Fisher process, but in reality birth is a
continuous process. The Wright-Fisher process corresponds to an underlying
diffusion process only in the limit that generation times are short. We
estimate the error introduced by using finite time intervals using the
uncensored data from the Netherlands (Fig~\ref{fig:names}b, Netherlands). When
aggregating uncensored Netherlands counts at 1 and 2 year intervals,
$\hat{s}(p)$ curves differ by at most 0.6\%/year (RMS error: 0.4) due to a
combination of bias and pseudoreplication noise.  Netherlands 1y and 5y
inferences differ by at most 1.3\%/year (RMS: 0.81, Fig~\ref{fig:names}b),
indicating the bias diminishes with finer sampling intervals. The bias
introduced by 1-yearly sampling is therefore less than 0.6\%/year in
Netherlands data, which we take to be representative of all name datasets. 

\subsection{American Kennel Club}

\begin{figure}[t]
\noindent
\centering
\includegraphics[width=120mm]{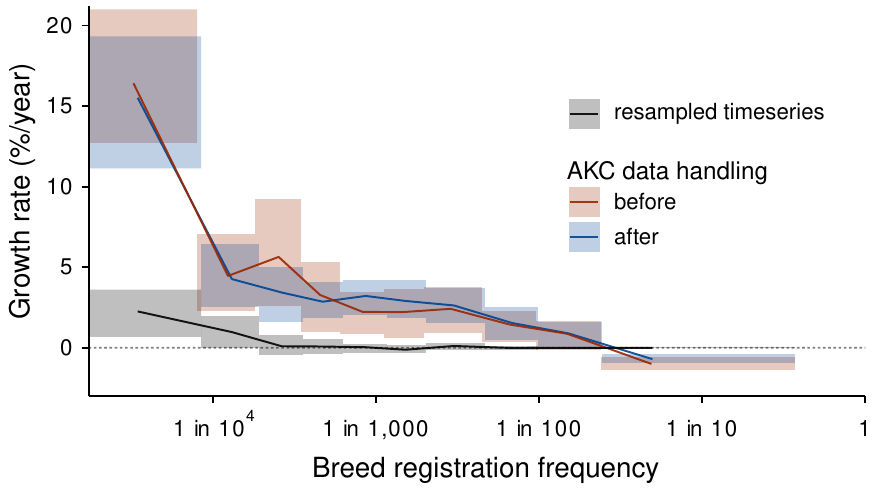}
\caption{\textbf{AKC data handling and sampling bias.} In the inference from
AKC data depicted in Fig.~\ref{fig:dogs} (blue here), we ignored transitions to
or from zero, removed outliers, removed data from the year of introduction of
dog breeds, and severed time series at apparent discontinuities
(\ref{sec:dogs}). Here we show the inference from the raw data (red) as well as
the inference from a time series of samples (black). The sampling bias inference
does not exceed 23\% proportional error on the magnitude of the inferred
selection coefficient (blue curve) or 2.3\%/year absolute magnitude, and
remains well within the statistical error in each bin.}
\label{fig:akcout}
\end{figure}

\label{sec:dogs}

We obtained from a public depository\cite{ghirlanda2013american,
ghirlanda2013fashion} an annual time-series of breed registrations from the
American Kennel Club (AKC), which includes a total of 53,697,706 dogs of 153
different breeds registered between the years 1926 and 2005.  After checking
data quality and removing outliers associated with known or imputed errors in
breed classification, we inferred frequency-dependent selection using the
procedure described in Section~\ref{sec:sp}. We also measured sampling bias as
described in Section~\ref{sec:samp}. In the lowest two bins, sampling bias was
appreciable but still less than statistical error, whereas sampling bias was
otherwise negligible. The net effect on the inference of sampling bias and data
handling including removing outliers, separating time series and ignoring
transitions to or from zero is depicted in Fig.~\ref{fig:akcout}.

We used counts of dog breeds as they appear in the original data with the
exception of counts $0$, which we omitted from the inference procedure because
of pervasive non-distinction between 0 and missing data in the original dataset.
That is, when $x_i = 0$ or $x'_i = 0$, we simply omit $x_i$ and $x'_i$ from any
sums in equations such as Eq.~\ref{eq:mmiterates}. Thus $m_t$ is always zero
(we do not infer mutation rate), $n_t = n' = \sum_{i=1}^{k_{t-1}} x_i^t$ where
$i$ ranges over the $k_{t-1}$ types which were present (nonzero) at timestep
$t-1$, and the sums over types in Eq.~\ref{eq:mmiterates} involve only the $k$
types which are present at $t$, ignoring the subpopulation of types for which
there is no data.  We divided the frequency range into 10 bins, choosing bin
boundaries by quantiles so that each bin incorporated a roughly equal share of
transitions $x_i \to x'_i$ (range: 848-894).

We excluded rows in the dataset (nominally, breeds) containing breed subtypes
that are summed in other rows, e.g.\ the rows ``Dachshund-XX'' and
``Dachshund-XX  (long-haired)'' are summed in ``Dachshund'', and so we retain
only ``Dachshund''.  We retained ``Dachshund'', ``English Toy Spaniel (All)'',
``Fox Terrier'', ``Manchester Terrier'', ``Poodle'' and ``Spaniel (All
Cockers)'' while excluding all of their subtype rows.

We also removed artifactual discontinuities from the time series. The
evolutionary models we fit have approximately continuous frequency trajectories
\cite{kimura1964diffusion}, whereas the AKC data contains sharp discontinuities
arising from reclassification of breeds or changing interpretation of
registration data over time. Hence we omitted certain counts from the data or
broke some time series into two separate time series (e.g. Greyhounds-pre1935 and
Greyhound-1935onward). We detected anomalous discontinuities by fitting the
frequency-dependent model to the raw data and examining residuals. We produced
a $p$-value for each residual transition in frequency as its quantile in the
Binomial distribution given its expected frequency and the current population
size. The most improbable transition (Chinese Shar-Pei 1992$\to$1993:
90,081$\to$19,465, Bonferroni-corrected $p$$<$$10^{-55}$) occurred immediately
after the Shar-Pei's year of introduction in the dataset, suggesting that the
count during the year of introduction differs in kind from subsequent counts.
In fact, 84 breeds were introduced over the course of the time series, and 40
out of these 84 have unlikely transitions ($p$$<$$0.1$) immediately following
their year of introduction (compared to 654 out of 8810 in the full data,
indicating otherwise slightly conservative $p$-values). We conclude that counts
are unreliable in the year a breed is first introduced into the dataset, and so
we ignore all 84 such counts.  Removing initial points from a time series does
not bias the inference under the frequency dependent model. 

We detected five other likely artifactual discontinuities by ranking
transitions according to the magnitude of proportional change. We removed the
count for Poodle in 1939 (effectively removing two transitions) because Poodle
is a sum of breed subtypes and the ``Poodle (miniature)'' subtype was undefined
in 1939.  We also split the time series for Manchester Terrier at 1945/1946
(where it nominally experienced a transition 880$\to$61), Curly-Coated
Retriever at 1932/1933 (82$\to$1), and Greyhound at 1934/1935 (12$\to$1624).
Manchester Terrier is a sum over two sub-types, and the end of the ``Manchester
Terrier (Toy)'' time series in 1945 causes the discontinuity.  We have no
evidence that the abrupt transitions for Greyhound and Curly-Coated Retriever
correspond to errors in the data beyond the fact that they are the two highest
proportional changes in the entire dataset, and they are much larger than
changes known to be artifacts of errors such as breed reclassification. 

\section{Novelty bias model fits}

\begin{figure}[t] 
\noindent
\centering
\includegraphics[width=120mm]{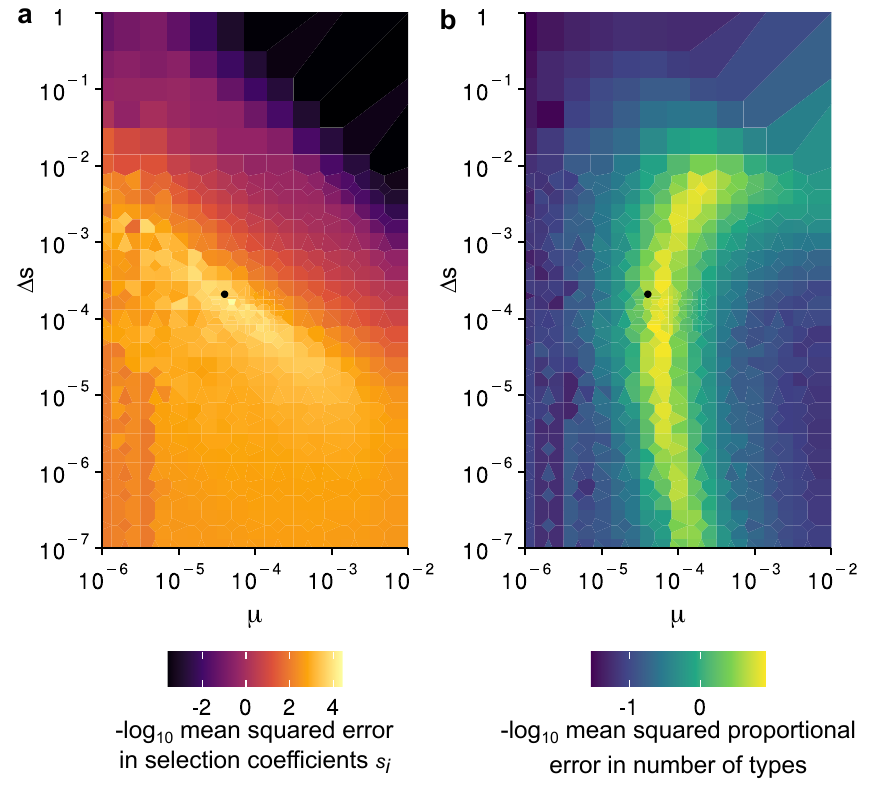}
\caption{\textbf{Novelty bias approximation to AKC depending on parameters.}
Inferred frequency-dependent selection (a) and type diversity (b) match the AKC
data for certain values for the parameters $\Delta s$, $\mu$ of the novelty
bias model. The black dot indicates the chosen parameters in
Fig.~\ref{fig:dogs}. Mean squared errors in selection coefficient (a) and
proportional number of types ($\log(n_{sim}/n_{AKC})^2$) (b) are computed by
averaging across bins and up to 12 replicate simulations.}
\label{fig:dognovgrid}
\end{figure}

\label{sec:dognov}

We compared the form of frequency dependent selection inferred from the AKC
time series to simulations of an evolutionary process that favors novel
types---a novelty bias model.  The novelty bias model is an infinite-alleles
Wright-Fisher process in a population of $N$ individuals with on average $N\mu$
new types introduced per generation.  We use four parameters: the constant
census population size $N$, the constant effective population size $N_e$, the
selection benefit to each novel type $\Delta s$, and the per capita per
generation mutation rate $\mu$.  We number each new type sequentially, giving
type number 1 fitness 1 (selection coefficient $s = 0$) and type number $n$
selection coefficient $s = n\Delta s$. Thus after long times, both the average
selection coefficient in the population (mean fitness) and the selection
coefficient of the fittest type increase at rate $N\mu\Delta s$ per generation
on average. Hence the relative fitness of the fittest type to the rest of the
population is also constant on average after long times. We allow the strength
of genetic drift to vary independently of census population size $N$ by
simulating $g$ timesteps of neutral evolution between each round of mutation
and selection, producing an effective population size $N_e = N/g$. Hence one
\textit{generation} of simulation is composed of one Wright-Fisher timestep
incorporating mutation and selection and $g - 1$ timesteps of neutral
Wright-Fisher evolution.

Notably, types in the novelty bias model are not exchangeable, as in our model
of pure frequency dependent selection, but nevertheless the novelty bias model
induces an effective frequency dependent growth rate---namely the average
growth rate of types that fall in a given frequency range.

We fitted parameters of the novelty bias model to the AKC data by conducting
grid searches across a range of $\mu$ ($10^{-7}$ to 0.005) and $\Delta s$
($10^{-7}$ to 0.5) on a roughly log evenly spaced grids
(Fig.~\ref{fig:dognovgrid}). We recorded the average selection coefficient in
simulations $s_\mathrm{ave}$ for types that fall in each frequency bin used in
the AKC inference---that is, we measured the equilibrium effective frequency
dependence in the novelty bias model.  We searched for parameter sets that
minimize the squared difference in the effective frequency dependence measured
in the simulations versus the frequency dependence inferred from the AKC data,
by summing over frequency bins $(s_\mathrm{ave} - \hat{s}_\mathrm{AKC})^2$. We
also looked for agreement in overall diversity of types by tracking the squared
difference in the log average number of types in each frequency range in the
simulation versus in the AKC data.  Parameter sets that agreed in diversity
were clustered in a narrow band of mutation rates; whereas parameter sets that
agreed in effective frequency dependence were clustered in a narrow diagonal
band with $\mu\Delta s \approx \mathrm{const.}$. At all values of $N$ and $N_e$
investigated, these two bands had an identifiable intersection that produced
good matches in both frequency dependence and diversity, indicating that $\mu$
and $\Delta s$ are identifiable given $N$ and $N_e$. Near the least-squares
minimum, the measured shape of effective frequency dependence from simulations
was similar for different values of $\Delta s$ holding other parameters
constant, so we easily matched the range of selection coefficients between the
simulation and AKC data (dominated by the growth rate of the rarest frequency
bin) by binary search on $\Delta s$.

We took $N$ to equal the maximum census population size in the AKC dataset in
order to cover the same range of frequencies, namely $N$=1,435,737.  Increasing
$N$ beyond this value and refitting $\mu$ and $\Delta s$ as above revealed
optimal parameter sets that lie on a  line of constant $N\mu\Delta s$. These
results concord with the intuition that the average rate of fitness increase
due to the appearance of novel types is what primarily determines the strength
of the effective frequency dependence. Varying $N_e$ showed that the particular
value of $N\mu\Delta s$ that coincided with the best fits was roughly
proportional to $N_e$, indicating that $N_e$ and $N\mu\Delta s$ are not
separately identifiable based on diversity and effective frequency dependence
alone.

Conversely, determining $N_e$ produced an absolute estimate of $N\mu\Delta
s$=1.2\% without knowledge of $N$.  Values of $N_e$ close to 10,000 and 50,000
produced stochastic fluctuations in novelty bias simulations that visually
match those in the AKC time series, and $N_e \approx 50,000$ gave the best
results for intermediate frequencies (Fig~\ref{fig:dogs}b vs c). The AKC
dataset itself displays different levels of stochasticity over time, as its
census population size varies over a factor of 30 from 46,788 in 1931 to
1,435,737 in 1992. To achieve this realistic $N_e$, we simulated with
$N$=1,435,737 and $g = 30$ giving an $N_e = 47,857.9$. Given these values of
$N$ and $N_e$, and following the fitting procedure described above, we chose
$\mu = 0.00004$ (to match the diversity of existing types observed in the AKC
time series) and we chose $\Delta s = 0.00021$ (to match the maximum $s$ in any
frequency bin in the AKC inference).  Since values of $N_e$ ranging from 15,000
to 150,000 are all plausible, and this range of $N_e$ far exceeds the error in
finding best fits of $\mu$ and $\Delta s$ given $N_e$, we estimated $N\mu\Delta
s$=0.012 up to roughly a factor of three in either direction.

\end{document}